\documentclass[%
 reprint,
 amsmath,amssymb,
 aps,
]{revtex4-2}

\usepackage{xcolor}
\usepackage{array}

\usepackage{caption}
\usepackage{subcaption}

\usepackage{graphicx}
\usepackage{dcolumn}
\usepackage{bm}
\usepackage{hyperref}
\usepackage{url}

\begin{document}

\preprint{APS/123-QED}

\title{Testing gravity with gauge-invariant polarization states of gravitational waves:\\Theory and pulsar timing sensitivity}

\author{Márcio E.S. Alves}
 \altaffiliation[Also at ]{Universidade Estadual Paulista (UNESP), Faculdade de Engenharia e Ciências, Departamento de Física e Química, Guaratinguetá, SP, 12516-410, Brazil.}
 \email{marcio.alves@unesp.br}
\affiliation{%
 Universidade Estadual Paulista (UNESP), Instituto de Ciência e Tecnologia,\\ São José dos Campos, SP, 12247-004, Brazil
}%

\date{\today}

\begin{abstract}
The determination of the polarization modes of gravitational waves (GWs) and their dispersion relations is a crucial task for scrutinizing the viability of extended theories of gravity. A tool to investigate the polarization states of GWs is the well-known formalism developed by Eardley, Lee, and Lightman (ELL) [Phys. Rev. D 8, 3308 (1973)] which uses the Newman-Penrose (NP) coefficients to determine the polarization content of GWs in metric theories of gravity. However, if the speed of GWs is smaller than the speed of light, the number of NP coefficients is greater than the number of polarizations. To overcome this inconvenience we use the Bardeen formalism to describe the six possible polarization modes of GWs considering general dispersion relations for the modes. The definition of a new gauge-invariant quantity enables an unambiguous description of the scalar longitudinal polarization mode. We apply the formalism to General Relativity, scalar-tensor theories, $f(R)$-gravity, and a wide class of quadratic gravity. To obtain a bridge between theory and experiment, we derive an explicit relation between a physical observable (the derivative of the frequency shift of an electromagnetic signal), and the gauge-invariant variables. From this relation, we find an analytical formula for the pulsar timing rms response to each polarization mode. To estimate the sensitivity of a single pulsar timing we focus on the case of a dispersion relation of a massive particle. The sensitivity curves of the scalar longitudinal and vector polarization modes change significantly depending on the value of the effective mass. The detection (or absence of detection) of the polarization modes using the pulsar timing technique has decisive implications for alternative theories of gravity. Finally, investigating a cutoff frequency in the pulsar timing band can lead to a more stringent bound on the graviton mass than that presented by ground-based interferometers.
\end{abstract}

\maketitle


\section{Introduction}

The gravitational wave (GW) events detected so far by the Advanced LIGO and Advanced Virgo interferometers have shown their ability to impact our knowledge of physics and astrophysics. These observations offer a unique opportunity to test General Relativity (GR) in the dynamical regime.

All the extensions to Einstein's theory predict modifications to the conventional GW signal due to one or more of three aspects, namely, changes in the waveform due to particularities in the generation mechanism, changes in the propagation due to new dispersion relations or differences in the interaction of the wave with the background geometry, and the number of independent polarization states of GWs.

Considering these effects, tests of gravity performed with the data of the three observing runs of Advanced LIGO and Advanced Virgo have shown that all the observed events are consistent with GR \cite{LIGO2019,LIGO2021,LIGO2022}. However, the planned increase in the sensitivity of the detectors, the new generations of interferometers, the pulsar timing technique and the future space-based GW detectors as LISA will be able to produce stringent tests to GR. 

In the case of the polarization states of GWs, a detection indicating the presence of a polarization mode beyond the usual plus and cross polarizations would imply a violation of Einstein's theory. In general, an alternative theory of gravity in four dimensions can predict up to six polarization modes of GWs, namely, two tensor, two vector, one scalar transversal, and one scalar longitudinal \cite{Eardley1973a,Eardley1973b}. 

To check the presence or absence of such modes in a specific theory is appropriate to consider the evaluation of gauge-invariant quantities to warrant that they are related to truly physical observables. The most common strategy is to use the formalism developed by Eardley, Lee, and Lightman (ELL) fifty years ago \cite{Eardley1973a,Eardley1973b}. Within the ELL framework, the irreducible parts of the linearized Riemann tensor are related to the polarization states of GWs. These components can be written on a tetrad basis to obtain the so-called Newman-Penrose (NP) coefficients.

The NP formalism \cite{Newman1962a,Newman1962b} is a powerful framework to describe the spacetime properties of a Lorentzian geometry in a coordinate-independent manner. It has many applications not only for GR but also in the scope of extended theories of gravity including the study of perturbed spacetimes in such theories (see, e.g., \cite{Suvorov2019,Svarc2023,Li2023}).  

In the ELL framework, the specific case of perturbations around a flat background is considered, resulting in two real and two complex NP coefficients describing the polarization states of GWs in any four-dimensional metric theory of gravity. The ELL formalism has been applied in the scope of several theories to reveal the polarization properties of GWs (see, e.g., \cite{dePaula2004,Alves2009,AlvesCQG2010,Hohmann2012,Myung2014,Alves2016,Sharif2017,Bertolami2018,Abedi2018,Mebarki2019,Toniato2019,Wagle2019,Haghshenas2020,Gogoi2020,Dong2022,Dongb2022}).

In recent years, however, some criticisms have been raised in the literature regarding the use of the original ELL formalism in theories that present one or more massive modes in the linearized regime \cite{Liang2017,Gong2018,Hou2018,Hyun2019}. This is the case of a huge class of alternative theories, including the massive version of the Brans-Dicke theory \cite{AlvesCQG2010}, Horndeski theory \cite{Hou2018} and $f(R)$-gravity \cite{Alves2009,Sharif2017,Liang2017}. The main criticism is related to the fact that if GWs travel at a speed different from the speed of light, then other NP coefficients, beyond the original four, would be non-null. Consequently, the original ELL formalism would be incomplete and could result in misleading conclusions for some theories. Thus, in this case, to have a complete description of the polarizations one needs new NP variables. In fact, to describe six polarizations there are not only four but nine NP coefficients representing fourteen components of the Riemann tensor \cite{Hyun2019}. Certainly, some coefficients could have a greater amplitude when compared to others depending on the dispersion relation and frequency.

However, there is a gauge-invariant alternate formalism to identify the polarization modes in a metric theory of gravity. It consists of decomposing the metric into irreducible components according to their properties under spatial rotation [(3+1) decomposition] and then constructing gauge-invariant combinations of the metric perturbations. In cosmological perturbation theory, such quantities are known as Bardeen variables \cite{Bardeen1980,Mukhanov1992}. Flanagan and Hughes \cite{Flanagan2005} have used these variables in GR to describe perturbations over a Minkowski background. Recently, Wagle et al. \cite{Wagle2019} used these gauge-invariant variables to study the polarizations of GWs in the context of two theories, namely, the dynamical Chern-Simons gravity and the Einstein-dilaton-Gauss-Bonnet gravity. They have found that the ELL formalism and the (3+1) decomposition lead to the same result in both cases. In the case of $f(R)$ gravity, Moretti et al. \cite{Moretti2019} have applied the gauge-invariant formalism to investigate the propagating degrees of freedom within this class of theories. Using the same formalism, Dong et al. \cite{Dongc2023} have shown that the number of polarization modes of GWs depends on the choice of parameters in the generalized Proca theory. Other works in the literature have used this framework to describe GWs in alternative theories of gravity (see, e.g., \cite{Gong2018a,Gong2018b,Tachinami2021}). The formalism of Bardeen variables has the advantage that the same number of variables are applicable to describe the polarization modes of GWs to any frequency. 

The aim of the present article is twofold. First, we review the Bardeen formalism and show how it applies in identifying the polarization modes of GWs for any metric theory of gravity. We discuss the advantages of this formalism when compared to the EEL formalism. The second aim is to estimate the pulsar timing sensitivities considering the gauge-invariant variables to evaluate the response to the polarization modes. 

Analyzing the pulsar timing sensitivity, Alves and Tinto have shown that this technique is significantly more sensitive to non-transverse polarizations than the usual plus and cross polarizations \cite{Alves2011}. To a lesser extent, the same effect appears in the LISA frequency band \cite{Tinto2010}. Other authors have noticed the enhancement in the sensitivity to non-transverse polarization modes \cite{Chamberlin2012,Blaut2012}. The effect appears if the wavelength of GWs is of the same order or smaller than the scale size of the detector \cite{Alves2011,Tinto2010}. Therefore, one does not expect this effect to appear in detectors operating within the long-wavelength limit as in ground-based interferometers.

In deriving the pulsar timing sensitivity, previous works consider a specific gauge choice  (one of the most common is the synchronous gauge). Although this is a usual procedure, the resulting sensitivity curve may not be appropriate to evaluate the detectability of the polarization modes in all alternative theories of gravity. This is because a given gauge is not necessarily applicable in all theories and residual gauge freedom may be present, resulting in misleading interpretations. Furthermore, the most important aspect is that since gauge-invariant quantities can be related to truly physical observables, detector sensitivity is expected to be derived in terms of them.

In this sense, here we derive in a complete gauge-invariant fashion the pulsar timing sensitivities to each polarization mode considering the dispersion relation of a massive particle. We show how the sensitivity can be written in terms of the Bardeen variables used to describe the polarization modes of GWs in a wide range of metric theories of gravity.

The article is organized as follows. In Section (\ref{sec 2 NP}) we give a short overview of the  ELL formalism. In Section (\ref{sec 3 Bardeen}) we describe the Bardeen formalism and show how it can be applied in describing the polarization modes of GWs for any theory of gravity. We apply the formalism to the case of GR, scalar-tensor theories of gravity,  $f(R)$-gravity, and a wide class of quadratic gravity.  The relation between the one-way response to GWs and the gauge-invariant variables, as well as the pulsar timing sensitivity, are obtained in Section (\ref{sec 4 PT}). Finally, we conclude the article with Section (\ref{sec 5 conclusion}). Throughout the article, we use the metric signature $(-,+,+,+)$ and units such that $c = \hbar = 1$ unless otherwise mentioned.

\section{An overview of the ELL formalism}\label{sec 2 NP}

In the original  ELL formalism for the determination of polarization modes of GWs, Eardley {\it et al.} \cite{Eardley1973a,Eardley1973b} considered GWs propagating in the $+z$ direction at the speed of light and defined a null complex tetrad $(\mathbf{k, l, m, \bar{m}})$. This tetrad is related to the Cartesian tetrad $(\mathbf{e}_t, \mathbf{e}_x, \mathbf{e}_y, \mathbf{e}_z)$ by
\begin{equation}
    \mathbf{k} = \frac{1}{\sqrt{2}}(\mathbf{e}_t + \mathbf{e}_z),
\end{equation}
\begin{equation}
    \mathbf{l} = \frac{1}{\sqrt{2}}(\mathbf{e}_t - \mathbf{e}_z),
\end{equation}
\begin{equation}
    \mathbf{m} = \frac{1}{\sqrt{2}}(\mathbf{e}_x + i\mathbf{e}_y),
\end{equation}
\begin{equation}
    \bar{\mathbf{m}} = \frac{1}{\sqrt{2}}(\mathbf{e}_x -i \mathbf{e}_y).
\end{equation}

It is easy to verify that the tetrad vectors obey the relations:
\begin{equation}
    -\mathbf{k}\cdot\mathbf{l} = \mathbf{m}\cdot\bar{\mathbf{m}} = 1,
\end{equation}
\begin{equation}
    \mathbf{k}\cdot\mathbf{m} = \mathbf{k}\cdot\bar{\mathbf{m}} = \mathbf{l}\cdot\mathbf{m} = \mathbf{l}\cdot\bar{\mathbf{m}} = 0.
\end{equation}

To denote components of tensors with respect to the null tetrad basis we use Roman subscripts, that is, $P_{abc\dots} \equiv P_{\alpha\beta\gamma\dots}a^\alpha b^\beta c^\gamma \dots$, where $(a, b, c,\dots )$ run over $(\mathbf{k, l, m, \bar{m}})$ and $(\alpha, \beta, \gamma, \dots)$ run over $(t, x, y, z)$.

The Riemann curvature tensor $R_{\alpha\beta\gamma\delta}$ can be split into the irreducible parts: the Weyl tensor, the traceless Ricci tensor and the curvature scalar, whose tetrad components can be named, respectively as $\Psi_A$, $\Phi_{AB}$ and $\Lambda$ following the notation of Newman and Penrose \cite{Newman1962a,Newman1962b}. In general, in a four-dimensional space, we have ten $\Psi$'s, nine $\Phi$'s, and one $\Lambda$ which are all algebraically independent. However, when we restrict ourselves to null plane waves, we find that the differential and algebraic properties of $R_{\alpha\beta\gamma\delta}$ reduce the number of independent components to six. In the above tetrad, we can choose the following quantities to represent these components \cite{Eardley1973a,Eardley1973b}
\begin{equation}
    \Psi_2 = - \frac{1}{6}R_{lklk},
\end{equation}
\begin{equation}
    \Psi_3 = - \frac{1}{2}R_{lkl\bar{k}},
\end{equation}
\begin{equation}
    \Psi_4 = - R_{l\bar{m}l\bar{m}},
\end{equation}
\begin{equation}
    \Phi_{22} = - R_{lml\bar{m}}.
\end{equation}

Notice that since $\Psi_3$ and $\Psi_4$ are complex, each represents two independent polarizations. For these six components, three are transverse to the direction of propagation, with two representing quadrupolar deformations [${\rm Re}(\Psi_4)$ and ${\rm Im}(\Psi_4)$] and one monopolar deformation ($\Phi_{22}$). Three modes are longitudinal, with one an axially symmetric stretching mode in the propagation direction ($\Psi_2$), and one quadrupolar mode in each one of the two orthogonal planes containing the propagation direction [${\rm Re}(\Psi_3)$ and ${\rm Im}(\Psi_3)$].  

The above formalism is still accurate if the speed of GWs is close to the speed of light. In fact, corrections to the  ELL formalism are of the order $O(\epsilon \delta\mathcal{R})$ to the case of a nearly null GW  \cite{Will2018}, where $\epsilon = (c/v_{\rm gw})^2 - 1$, $v_{\rm gw}$ is the speed of GWs and $\delta \mathcal{R}$ is some component of the perturbed Riemann tensor. Therefore, considering the current upper bound for the graviton mass from the observations of binary black hole mergers ($m_g \leq 1.27 \times 10^{-23}~{\rm eV}/c^2$)  \cite{LIGO2022}, we find $\epsilon \sim 10^{-21}$ for the frequency 0.1 kHz (considering the dispersion relation of a massive graviton). For the frequency of 1 mHz, we obtain $\epsilon \sim 10^{-11}$. Thus, in these cases $O(\epsilon \mathcal{R})$ is several orders of magnitude smaller than the NP quantities for null waves [which are of the order $O(\mathcal{R})$ ]. We conclude that such corrections are undetectable in the frequency band of ground-based and space-based interferometers. 

On the other hand, they can become important for lower frequencies. For instance, in the band of pulsar timing arrays, we find $\epsilon \sim 1$ for the above mass of the graviton, and by considering frequencies of the order of nanohertz.  In this case, to have a complete description of the polarizations the ELL formalism needs to be amended by including other NP coefficients. Considering plane waves propagating at a speed $v_{\rm gw}$, Hyun {\it et al.} \cite{Hyun2019} have expressed the polarizations of GWs in terms of nine NP scalars, namely, $\Psi_0$, $\Psi_1$, $\Psi_2$, $\Psi_3$, $\Psi_4$, $\Phi_{00}$, $\Phi_{02}$, $\Phi_{22}$, $\Lambda$. Since $\Psi_0$, $\Psi_1$, $\Psi_3$, $\Psi_4$, $\Phi_{02}$ are complex, the nine scalars represent fourteen components of the Riemann curvature tensor needed to describe six polarization modes of GWs. One can use the formalism of Bardeen variables described in the next section to overcome this inconvenience.

\section{Describing the polarization states of GWs within the Bardeen framework}\label{sec 3 Bardeen}

\subsection{Helicity decomposition and gauge invariant perturbations}

In this section, we introduce the helicity decomposition of the metric perturbation and define the gauge invariant variables that can be computed from them.   

Let us start by expanding the metric around flat space $g_{\mu\nu} = \eta_{\mu\nu} + h_{\mu\nu}$ with $|h_{\mu\nu}|\ll 1$. The perturbation $h_{\mu\nu}$ can be decomposed considering the behavior of its components under spatial rotations as follows
\begin{equation}\label{eq h00}
     h_{00} = 2\psi,    
\end{equation}
\begin{equation}\label{eq h0i}
    h_{0i} = \beta_i + \partial_i \gamma,
\end{equation}
\begin{equation}\label{eq hij}
    h_{ij} = -2\phi \delta_{ij} + \left(\partial_i \partial_j -\frac{1}{3}\delta_{ij} \nabla^2 \right) \lambda + \frac{1}{2} (\partial_i \epsilon_j + \partial_j \epsilon_i) + h_{ij}^{\rm TT},
\end{equation}
where the vector and tensor quantities are subject to the following constraints
\begin{equation}
    \partial_i\beta^i = 0,~~\partial_i\epsilon^i = 0,
\end{equation}
\begin{equation}
    \partial^j h_{ij}^{\rm TT} = 0,~~\delta^{ij}h_{ij}^{\rm TT} = 0.
\end{equation}

Therefore, from the 10 degrees of freedom of $h_{\mu\nu}$ we have four scalar degrees of freedom $\{\psi, \phi, \gamma, \lambda\}$, four degrees of freedom in the two transverse vectors $\{\beta_i, \epsilon_i\}$ and two degrees of freedom in the transverse-traceless (TT) spatial tensor $h^{\rm TT}_{ij}$. 

The gauge transformation of the metric perturbation 
\begin{equation} \label{gauge transf}
    h_{\mu\nu} \rightarrow h_{\mu\nu} - (\partial_\mu \xi_\nu + \partial_\nu \xi_\mu ), 
\end{equation}
with $|\partial_\mu \xi_\nu|$ small preserves $|h_{\mu\nu}|\ll 1$, thus it is a symmetry of the linearized theory in general. To understand how the harmonic variables behave under this gauge transformation, notice that the 4-vector $\xi_\mu$ can also be decomposed as
\begin{equation}
    \xi_0 = A,
\end{equation}
\begin{equation}
    \xi_i = B_i + \partial_i C,
\end{equation}
with 
\begin{equation}
    \partial_i B^i = 0.
\end{equation}

Therefore, considering Eq. (\ref{gauge transf}) and following the symmetry of the transformations under spatial rotations, we find that the gauge transformations of the scalar harmonic variables read
\begin{align}\label{scalar transf}
    \psi  &\rightarrow \psi - \dot{A},  \nonumber \\
    \phi &\rightarrow \phi + \frac{1}{3} \nabla^2C, \nonumber \\
    \lambda &\rightarrow \lambda -2C, \nonumber \\
    \gamma  & \rightarrow \gamma - \dot{C} - A. 
\end{align}

The transformations of the vectors are
\begin{align}\label{vector transf}
    \beta_i & \rightarrow \beta_i - \dot{B_i}, \nonumber \\
    \epsilon_i & \rightarrow \epsilon_i - 2B_i,
\end{align}
while $h_{ij}^{TT}$ is gauge-invariant
\begin{align}
    h_{ij}^{TT} \rightarrow h_{ij}^{TT}.
\end{align}

The Riemann tensor and, correspondingly, the Einstein tensor are gauge invariant quantities. Therefore, one possible way of dealing with this gauge freedom is to impose gauge conditions on the metric perturbations. This is the usual way in GW physics. Several gauges are possible, some of the most common gauge choices are the synchronous gauge and the Newtonian gauge. The latter fixes the gauge uniquely, however, the conditions imposed by the former leave a residual gauge freedom. This ambiguity implies the existence of unphysical modes when the gravitational equations are solved. Particularly, this can lead to an ambiguity in the determination of truly propagating GW modes in alternative theories of gravity. In the Bardeen words ``{\it only gauge-invariant quantities have any inherent physical meaning}'' \cite{Bardeen1980}. 

In this sense, Bardeen \cite{Bardeen1980} has constructed gauge-invariant quantities from combinations of the above scalar and vector variables to deal with perturbations in a FLRW background spacetime. In this article, we consider solely the Minkowski metric for the background. From the transformations (\ref{scalar transf}) we see that we can obtain the following two gauge-invariant scalar combinations 
\begin{align}
    \Phi & = - \phi - \frac{1}{6} \nabla^2 \lambda, \\
    \Psi & = - \psi + \dot{\gamma} - \frac{1}{2} \ddot{\lambda}.
\end{align}

In the same way, from the transformations (\ref{vector transf}) we obtain one gauge-invariant transverse spatial vector
\begin{equation}
    \Xi_i = \beta_i - \frac{1}{2} \dot{\epsilon}_i,~~\partial_i \Xi^i = 0.
\end{equation}

Thus, we have six gauge-invariant degrees of freedom: two scalars, $\Psi$ and $\Phi$, two degrees of freedom in the spatial vector $\Xi_i$ and two degrees of freedom in the transverse-traceless spatial tensor $h_{ij}^{TT}$. These gauge-invariant variables are the flat background version of the well-known Bardeen variables. 

We can now write the electric components of the first order perturbed Riemann tensor using the (3+1) decomposition of the metric perturbations
\begin{equation}\label{Riemann 1st}
    \delta R_{i0j0} = \partial_i\partial_j \Psi - \ddot{\Phi} \delta_{ij} + \frac{1}{2}(\partial_i \dot{\Xi}_j + \partial_j \dot{\Xi}_i) - \frac{1}{2} \ddot{h}_{ij}^{\rm TT}.
\end{equation}
As expected, the perturbed Riemann tensor depends only on the gauge-invariant variables $\Phi$, $\Psi$, $\Xi_i$, and $h_{ij}^{\rm TT}$. It will also be useful to know the components of the perturbed Ricci tensor
\begin{align}
    \delta R_{00} & = \nabla^2\Psi - 3\ddot{\Phi}, \\
    \delta R_{0i} & = -2 \partial_i\dot{\Phi} - \frac{1}{2} \nabla^2 \Xi_i, \\
    \delta R_{ij} & = -\partial_i\partial_j(\Phi + \Psi) - \delta_{ij} (-\ddot{\Phi} + \nabla^2 \Phi) \nonumber \\
     & - \frac{1}{2}(\partial_i \dot{\Xi}_j  + \partial_j \dot{\Xi}_i) - \frac{1}{2} \Box h_{ij}^{\rm TT},
\end{align}
and the perturbed curvature scalar
\begin{equation} \label{curvature scalar}
    \delta R = - 2\nabla^2\Psi + 6\ddot{\Phi} - 4\nabla^2\Phi.
\end{equation}

From the above expressions we find the components of the perturbed Einstein tensor
\begin{equation}\label{Components of the Einstein tensor}
   \delta G_{00} = - 2 \nabla^2 \Phi,~~~\delta G_{0i} = \delta R_{0i},
\end{equation}    
\begin{equation}\label{ij component of the Einstein tensor}
   \delta G_{ij} = \delta R_{ij} - \frac{1}{2}\delta_{ij}\delta R.
\end{equation}

If $\dot{\Psi} \neq 0$ it will also prove useful to define the following gauge-invariant variable
\begin{equation}\label{Theta def} 
    \Theta \equiv \eta_{\Psi}^2 \Psi - \Phi,
\end{equation}
where 
\begin{equation}\label{eta def}
    \eta_\Psi \equiv \left|\frac{\nabla \Psi}{\dot{\Psi}}\right|.
\end{equation}

The physical meaning of $\Theta$ will be clarified in the next section. 

\subsection{Description of the polarization states of GWs with Bardeen variables\label{Description of pol with Bardeen var}}

The six degrees of freedom encoded in the four gauge-invariant variables defined above can be radiative or non-radiative depending on the underlying theory of gravity. It is well known that the transverse-traceless tensor $h_{ij}^{\rm TT}$ is the only radiative quantity in the GR theory. Moreover, for those theories of gravity predicting spin-0 polarization modes, it is expected that the scalars $\Phi$ and $\Psi$ are related through the field equations. This issue will be analyzed in the next section through examples in the context of some theories of gravity. In the present section, we describe in a general way the six polarization states of GWs employing the Bardeen variables. To this end, we suppose that all gauge-invariant quantities are radiative and independent. 

Since the variables are radiative, they are functions of the retarded time
\begin{equation}
    u \equiv t - \frac{\vec{k}\cdot\vec{r}}{\omega},
\end{equation}
where $\vec{k}$ is the GW wave vector and $\omega$ is the angular frequency. Consider, for instance, the scalar $\Psi = \Psi(u)$. In terms of the coordinates $t$ and $\vec{r}$ it obeys
\begin{equation}
    \nabla \Psi = -\frac{\vec{k}_\Psi}{\omega} \dot{\Psi},
\end{equation}
where $\vec{k}_\Psi$ is the wave vector of the $\Psi$ variable.

Using this equation in the definition (\ref{eta def}) we identify $\eta_\Psi$ as the dispersion relation
\begin{equation}
    \eta_\Psi(\omega) = \frac{k_\Psi(\omega)}{\omega},
\end{equation}
with $k_\Psi = |\vec{k}_\Psi|$.

Since each variable can have a different dispersion relation, we express them as functions of four different retarded times $u_A = t - \eta_A \hat{k}\cdot \vec{r}$, and the four dispersion relations are expressed by the functions $\eta_A (\omega) = k_A(\omega)/\omega$ with $A = \Phi$, $\Psi$, $V$ and $T$.

Hence, using the definition of the variable $\Theta$ (\ref{Theta def}) the electric components of the perturbed Riemann tensor (\ref{Riemann 1st}) now read
\begin{align}\label{Riemann 3rd}
    \delta R_{0i0j} &=  \hat{k}_i\hat{k}_j\Theta^{\prime\prime} - (\delta_{ij} - \hat{k}_i\hat{k}_j) \Phi^{\prime\prime} \nonumber \\ 
    &- \frac{1}{2}\eta_V(\hat{k}_i\Xi_j^{\prime\prime} + \hat{k}_j\Xi_i^{\prime\prime}) - \frac{1}{2}h_{ij}^{{\rm TT} {\prime\prime}},
\end{align}
where primes denote derivatives with respect to the retarded times and $\hat{k}_i$ are components of the unit wave vector. The transverse conditions can be written as $\hat{k}^i\Xi_i = 0$ and $\hat{k}^ih^{\rm TT}_{ij} = 0$. The two tensor polarization states are described by $h_{ij}^{\rm TT}$ as usual. The two spin-1 polarization states are described by the vector $\Xi_i$. Although $\Xi_i$ is transverse to the direction of propagation of the GW, notice that it enters $\delta R_{0i0j}$ with a term that is proportional to $\hat{k}_i$, which comes from the spatial derivative of $\Xi_i$. Therefore, we arrive at the known result that vector polarization affects the curvature in the transverse and longitudinal directions. 

In the second term in the right-hand-side of Eq. (\ref{Riemann 3rd}), we recognize the quantity multiplying the scalar Bardeen variable $\Phi$ as the projection operator $P_{ij} = (\delta_{ij} - \hat{k}_i\hat{k}_j)$ which has the property $\hat{k}^iP_{ij} = 0$. This operator projects any spatial vector on the subspace orthogonal to the direction of propagation of the GW. Thus, this term represents the scalar-transverse polarization mode. Finally, the term $\hat{k}_i\hat{k}_j\Theta^{\prime\prime}$ expresses the overall longitudinal effect from the scalar variables and then $\Theta$ is the degree of freedom responsible for describing the scalar-longitudinal polarization mode.

To summarize, the six possible polarization modes of GWs can be described by the six degrees of freedom present in the gauge-invariant variables $h_{ij}^{\rm TT}$, $\Xi_i$, $\Phi$, and $\Theta$. Since this result is valid for any dispersion relation, it turns out that the Bardeen formalism is much simpler than the ELL formalism in the determination of the polarization modes. 

\subsection{Lorentz transformations of the gauge-invariant variables}\label{sec: Lorentz transf}

Although the Bardeen variables are gauge-invariant quantities, generally they are not Lorentz-invariant. This means that distinct observers connected by boosts can measure different polarization contents of a given theory of gravity.

A finite Lorentz transformation can be built up from a sequence of infinitesimal Lorentz transformations. Therefore, if a quantity is invariant under infinitesimal boosts, it is also invariant under finite boosts. Thus, to evaluate the behavior of the gauge-invariant variables under boosts let us consider an infinitesimal Lorentz transformation $x^\mu \rightarrow x^{\prime\mu} = {\Lambda^\mu}_\nu x^\nu$ with ${\Lambda^\mu}_\nu = {\delta^\mu}_\nu + {\omega^\mu}_\nu$. The metric perturbation changes according to

\begin{equation}
    h_{\mu\nu}^\prime = h_{\mu\nu} + \delta h_{\mu\nu},
\end{equation}
where
\begin{equation}
    \delta h_{\mu\nu} = {\omega_\mu}^\rho h_{\rho\nu} + {\omega_\nu}^\rho h_{\mu \rho}.
\end{equation}

The non-null components of the quantity $\delta h_{\mu\nu}$ are small compared with $h_{\mu\nu}$, however when considering a sequence of infinitesimal Lorentz transformations the transformed $h^\prime_{\mu\nu}$ can be significantly different from $h_{\mu\nu}$.

Using this transformation and remembering the decomposition of the metric perturbation in terms of the harmonic variables given by Eqs. (\ref{eq h00}), (\ref{eq h0i}) and (\ref{eq hij}) one can find the change of each variable due to infinitesimal Lorentz transformations. If one restricts to boosts $\omega_{ij} = 0$ and we have the following expressions for the change in the gauge-invariant variables \cite{Jaccard2013}\footnote{Notice that our definition of the variable $\Psi$ differs from the definition given by the Ref. \cite{Jaccard2013} by a minus sign, i.e., one should change $\Psi \rightarrow -\Psi$ to compare the equations presented in both articles.}
\begin{equation}\label{eq Lorentz phi}
    \delta \Phi = \frac{1}{2}{\omega_0}^i \Xi_i,
\end{equation}
\begin{equation}\label{eq Lorentz Psi}
    \delta \Psi = - 2{\omega_0}^i\nabla^{-2}\partial_i(\dot{\Phi}-\dot{\Psi})+ {\omega_0}^i \Xi_i - \frac{3}{2}{\omega_0}^i\nabla^{-2} \ddot{\Xi_i},
\end{equation}
\begin{align}\label{eq Lorentz Xi}
    \delta\Xi_i &= {\omega_0}^j \nabla^{-2}{\Big [}\Box h_{ij}^{TT} - (\partial_i\dot{\Xi}_j + \partial_j\dot{\Xi}_i) \nonumber \\
    & - 2(\partial_i\partial_j - \delta_{ij}\nabla^2)(\Phi - \Psi) {\Big ]}, 
\end{align}
\begin{align}\label{eq Lorentz h}
    \delta h_{ij}^{TT} &= \omega_{0i}\Xi_j + \omega_{0j}\Xi_i - \delta_{ij}{\omega_0}^k\Xi_k \nonumber \\
    & + {\omega_0}^k\nabla^{-2}{\Big [}\partial_i\partial_j\Xi_k - \partial_i\partial_k\Xi_j - \partial_j\partial_k\Xi_i \nonumber \\
    & - (\partial_i\dot{h}_{jk}^{TT}+\partial_j\dot{h}_{ik}^{TT}){\Big ]}.
\end{align}

Therefore, a consequence of the decomposition scheme is that the Bardeen variables transform among themselves under boosts.

As we will show in the next section, analyzing the linearized vacuum field equations of a theory of gravity, one can find the governing equations for the Bardeen variables and possible relations between the scalars $\Phi$ and $\Psi$. Although the variables are not Lorentz invariant, in general, it is interesting to notice some particular cases:
\begin{enumerate}
    \item $h_{ij}^{TT} \neq 0,~\Box h_{ij}^{TT} = 0,~\Xi_i = 0,~\Phi = \Psi$. The variables $\Phi$, $\Psi$ and $\Xi_i$ are Lorentz invariant.  
    \item $h_{ij}^{TT} \neq 0,~\Box h_{ij}^{TT} \neq 0,~\Xi_i = 0,~\Phi = \Psi$. The variables $\Phi$ and $\Psi$ are Lorentz invariant. 
    \item $h_{ij}^{TT} \neq 0,~\Xi_i = 0,~\Phi \neq \Psi$. The variable $\Phi$ is the only Lorentz invariant. 
\end{enumerate}

As a consequence of cases 1 and 2, the variable $\Theta$ is also Lorentz invariant, and thus all the Lorentz observers measure the same scalar-longitudinal mode. The scalar-transversal polarization mode is Lorentz invariant in the three cases. On the other hand, the vector polarization mode is invariant only in the special case 1 for which it is null for all Lorentz observers. If $\Xi_i \neq 0$ and/or $h_{ij}^{TT} \neq 0$, the vector and tensor modes are not Lorentz invariant in general.

\subsection{Polarization states of GWs in some theories of gravity}\label{subsec: polarizations in alternatives} 

The procedure of determination of the polarization modes in an alternative theory of gravity starts, as usual, with the linearization of the vacuum field equations of the theory. The equations for metric perturbations should be written in terms of the Bardeen variables. Finally, from these equations, it will be possible to determine which variable represents a truly radiative mode, the number of independent degrees of freedom, and each dispersion relation related to the polarization modes. In this section, we illustrate this procedure by evaluating the polarization modes of GWs for General Relativity, scalar-tensor theories of gravity, and $f(R)$-gravity. 

\subsubsection{General Relativity}

Let us consider the Einstein-Hilbert action in the absence of matter sources with a vanishing cosmological constant
\begin{equation}
    I = \frac{1}{16\pi G}\int d^4x \sqrt{-g} R.
\end{equation}

The Einstein equations for vacuum is obtained if we vary this action with respect to the metric $g_{\mu\nu}$. They are
\begin{equation}\label{Einstein eqs vacuum}
    G_{\mu\nu} = 0.
\end{equation}

Let us expand the metric about the Minkowski spacetime $g_{\mu\nu} = \eta_{\mu\nu} + h_{\mu\nu}$ and write the Einstein equations to first order in $h_{\mu\nu}$. If we use the gauge-invariant quantities described previously we obtain the components of the Einstein tensor as given by the Eqs. (\ref{Components of the Einstein tensor}) and (\ref{ij component of the Einstein tensor}). From the 00 component, we obtain
\begin{equation}\label{eq. for phi GR}
    \nabla^2 \Phi = 0.
\end{equation}

With this result and using Eq. (\ref{curvature scalar}) in the trace equation $g^{\mu\nu}G_{\mu\nu} = - R = 0$ we have 
\begin{equation}\label{eq for psi GR}
    \ddot{\Phi} - \frac{1}{3}\nabla^2\Psi = 0.    
\end{equation}

From Eq. (\ref{eq. for phi GR}) we conclude that in the absence of matter 
\begin{equation}
    \Phi = 0,
\end{equation}
and with this result in the Eq. (\ref{eq for psi GR}) we have $\nabla^2 \Psi = 0$ and then 
\begin{equation}
    \Psi = 0.
\end{equation}

Therefore, the two gauge-invariant scalars are non-propagating degrees of freedom in Einstein theory and vanish in the absence of matter fields. If we use this result in the $0i$ components of the Eq. (\ref{Einstein eqs vacuum}) along with Eq. (\ref{Components of the Einstein tensor}) we obtain a similar result for the vector modes
\begin{equation}
    \nabla^2 \Xi_i = 0~\Rightarrow~\Xi_i =  0.
\end{equation}

Applying the above results and Eq. (\ref{ij component of the Einstein tensor}) in the spatial components of the field equations $G_{ij} = 0$ we find the following equation for gauge-invariant tensor perturbation
\begin{equation}
    \square h_{ij}^{TT} = 0.
\end{equation}

Therefore, we see that only the tensor degrees of freedom are radiative since they obey a wave equation. In the absence of matter $\Phi = \Psi = 0$ and $\Xi_i = 0$, we have only two polarization states represented by $h_{ij}^{TT}$, which are the usual $+$ and $\times$ polarizations. Since the tensor modes propagate at the speed of light we have $\eta_T = 1$ for GR. From the Lorentz transformations of the Bardeen variables given by Eqs. (\ref{eq Lorentz phi}), (\ref{eq Lorentz Psi}), (\ref{eq Lorentz Xi}) and (\ref{eq Lorentz h}), we see that the absence of scalar and vector polarization modes is a Lorentz invariant statement in the GR case.

\subsubsection{Scalar-tensor theories of gravity}\label{subsection ST theory}

For simplicity, we restrict to scalar-tensor theories of gravity whose action can be written in the following form in the absence of matter \cite{Wagoner1970,Bergmann1968}
\begin{equation}\label{scalar field action}
    I = \frac{1}{16\pi}\int d^4x \sqrt{-g} \left[\varphi R - \frac{\varpi(\varphi)}{\varphi}\nabla^\mu\varphi\nabla_\mu\varphi + V(\varphi) \right]. 
\end{equation}

However, as will be clear at the end of this subsection, the results we will find are valid for more general scenarios encompassed by the Horndeski theory.

In the theory described by (\ref{scalar field action}), gravity is mediated not only by the metric but also by a scalar field $\varphi$, $\varpi(\varphi)$ is a coupling function and $V(\varphi)$ is a generic scalar field potential. Varying this action with respect to the metric and scalar field we obtain 
\begin{align}\label{Field eqs scalar-tensor}
    G_{\mu\nu} &= \frac{1}{2}\varphi^{-1}V(\varphi)g_{\mu\nu} + \varpi(\varphi)\varphi^{-2}{\Big ( }\nabla_\mu\varphi\nabla_\nu\varphi \nonumber \\
    & - \frac{1}{2}g_{\mu\nu}\nabla_\alpha\varphi\nabla^\alpha\varphi{\Big )} + \varphi^{-1} (\nabla_\mu\nabla_\nu\varphi - g_{\mu\nu}\square\varphi),
\end{align}
and 
\begin{equation}\label{field eq scalar field}
    \square\varphi + \frac{\varphi V^\prime(\varphi) - 2 V(\varphi)}{3 + 2\varpi(\varphi)} = - \frac{\varpi^\prime(\varphi)\nabla_\alpha\varphi\nabla^\alpha\varphi}{3 + 2\varpi(\varphi)},
\end{equation}
where a prime denotes derivative with respect to $\varphi$. 

In the weak-field limit, we can expand the metric about the Minkowski background as in the GR case, while the scalar field is expanded as
\begin{equation}
    \varphi = \varphi_0 + \delta \varphi,~~\delta \varphi \ll \varphi_0,
\end{equation}
where $\varphi_0$  is the asymptotic value of $\varphi$ far from the system that is generating GWs. Expanding the potential and the coupling function about $\varphi_0$ up to the second order we have
\begin{align}
    V(\varphi) &= V(\varphi_0) + V^\prime(\varphi_0)\delta\varphi + \frac{1}{2}V^{\prime\prime}(\varphi_0) \delta\varphi^2 + {\mathcal{O}}(\delta \varphi^3),\\
    \varpi(\varphi) &= \varpi(\varphi_0) + \varpi^\prime(\varphi_0)\delta\varphi + \frac{1}{2}\varpi^{\prime\prime}(\varphi_0) \delta\varphi^2 + {\mathcal{O}}(\delta \varphi^3).
\end{align}

Far from the system we assume that the spacetime is asymptotically Minkowski. Thus, imposing that the background Einstein tensor vanishes in the Eq. (\ref{Field eqs scalar-tensor}), we are lead to $V(\varphi_0) = V^\prime(\varphi_0) = 0$. In this limit, from (\ref{field eq scalar field}) we obtain the equation for the first-order perturbation of the scalar field
\begin{equation}\label{linear eq for scalar field}
    \left(\square - m^2\right)\delta \varphi = 0,
\end{equation}
where now $\square$ is the D'Alembertian operator for the Minkowski background metric and the mass of the scalar field is defined by 
\begin{equation}
    m^2 \equiv - \frac{\varphi_0 V^{\prime\prime}(\varphi_0)}{3 + 2\varpi_0},
\end{equation}
where $\varpi_0 = \varpi(\varphi_0)$. The first-order perturbation of Eqs. (\ref{Field eqs scalar-tensor}) become
\begin{equation}\label{linearized eqs scalar tensor}
    \delta G_{\mu\nu} + \left( \eta_{\mu\nu} \square - \partial_\mu\partial_\nu\right) \frac{\delta \varphi}{\varphi_0} = 0,
\end{equation}
where $\delta G_{\mu\nu}$ is the linearized Einstein tensor. Replacing the Eq. (\ref{Components of the Einstein tensor}) in the 00 component of this equation we find
\begin{equation}
    \nabla^2 \left( \Phi + \frac{1}{2}\frac{\delta \varphi}{\varphi_0} \right) = 0.
\end{equation}

Hence, in the absence of matter, we conclude that the gauge-invariant scalar $\Phi$ is proportional to the scalar field perturbation
\begin{equation}
    \Phi = - \frac{1}{2}\frac{\delta \varphi}{\varphi_0}.
\end{equation}

Moreover, using the trace of Eqs. (\ref{linearized eqs scalar tensor}), the Eq. (\ref{linear eq for scalar field}) and the perturbed expression of the Ricci scalar (\ref{curvature scalar}), it is easy to show that
\begin{equation}\label{eq for phi ST theory}
    \Psi = \Phi = - \frac{1}{2}\frac{\delta \varphi}{\varphi_0}.
\end{equation}

The $0i$ components of Eqs. (\ref{linearized eqs scalar tensor}) together with Eq. (\ref{Components of the Einstein tensor}) lead to the following equation for the gauge-invariant vector perturbation
\begin{equation}
    \nabla^2\Xi_i = 0,
\end{equation}
which, in the absence of matter, gives
\begin{equation}\label{eq for xi ST theory}
    \Xi_i = 0.
\end{equation}

Finally, using Eqs. (\ref{ij component of the Einstein tensor}), (\ref{eq for phi ST theory}) and (\ref{eq for xi ST theory}), the spatial components of Eqs. (\ref{linearized eqs scalar tensor}) result
\begin{equation}\label{tensor pert ST theory}
    \square h_{ij}^{TT} = 0.
\end{equation}

Therefore, we conclude that in the scalar-tensor theory of gravity, there are three radiative degrees of freedom. The two tensor degrees of freedom obey the wave equation (\ref{tensor pert ST theory}) in the same manner as in the Einstein theory. They propagate at the speed of light and, therefore, $\eta_T = 1$. On the other hand, there is a scalar degree of freedom $\Phi = \Psi$ which obeys a Klein-Gordon type equation 
\begin{equation}
    \left(\square - m^2\right)\Phi = 0.
\end{equation}

This equation has a solution $\Phi \propto { e}^{ik_\alpha x^\alpha}$ with the wave 4-vector $k^\alpha \equiv (\omega, \vec{k})$ respecting the dispersion relation 
\begin{equation}
    \omega^2 = k^2 + m^2,
\end{equation}
and, therefore, the function $\eta_A(\omega) = k_A(\omega)/\omega$ is given by
\begin{equation}
    \eta_{\Phi}(\omega) = \eta_{\Psi}(\omega) = \sqrt{1 - \left(\frac{m}{\omega}\right)^2}.
\end{equation}

Notice that this is a propagating mode provided that $\omega > m$. Hence, $m$ is a cutoff frequency for the massive scalar degree of freedom. 

Evaluating the scalar longitudinal gauge-invariant variable defined by Eq. (\ref{Theta def}) we find
\begin{equation}\label{eq: relation theta and mass}
    \Theta = -\left(\frac{m}{\omega} \right)^2\Phi.
\end{equation}

Thus, there is a non-zero contribution to the Riemann tensor in the direction of propagation of the scalar GW. The meaning of this result is that although one has only one scalar degree of freedom, it generates the effects of the two scalar polarization states, the scalar transversal and the scalar longitudinal modes. If $m = 0 \Rightarrow \Theta = 0$, the longitudinal effect vanishes and one restores the result of the original massless Brans-Dicke theory for which there is only the scalar transversal polarization mode \cite{Eardley1973a,Eardley1973b}.

Notice that GWs in the scalar-tensor theories of gravity enter in case 1 discussed in Section \ref{sec: Lorentz transf}. Therefore the variables $\Phi$, $\Theta$ and $\Xi_i$ are Lorentz invariant quantities. 

Although in the present derivation we have considered the action (\ref{scalar field action}), the results are valid for the Horndeski theory \cite{Horndeski1974}, which is the most general scalar-tensor theory of gravity with second-order equations of motion. This is because our results depend essentially on the weak field equations (\ref{linear eq for scalar field}) and (\ref{linearized eqs scalar tensor}). The linearized field equations of Horndeski theory acquire exactly these forms, with a redefinition of the mass $m$ (see Eqs. (17) and (18) of the Ref. \cite{Hou2018}). In Ref. \cite{Hou2018}, there is a similar discussion about the polarization states in scalar-tensor theories, though a different approach has been used. However, if the Palatini formalism is used to study the Horndeski theory, the number of scalar polarization modes can vary depending on the chosen parameters \cite{Dong2022}.

For purposes that will be clear in the following sections, notice that from the trace of Eq (\ref{linearized eqs scalar tensor}) and Eq. (\ref{linear eq for scalar field}) we obtain $\delta R = 3m^2 \delta \varphi/\varphi_0$. Thus, provided $m\neq 0$ the linearized field equations in scalar-tensor theory can be regarded as
\begin{equation}\label{eq perturbed eq scalar tensor}
    \delta G_{\mu\nu} + \frac{1}{3m^2}\left(\eta_{\mu\nu}\Box - \partial_\mu \partial_\nu \right) \delta R = 0, 
\end{equation}
with trace given by
\begin{equation}
    \left( \Box - m^2 \right)\delta R = 0.
\end{equation}

In this way, the scalar field does not appear explicitly in the equations except in the definition of the mass.

\subsubsection{$f(R)$-gravity}

The action for the $f(R)$-gravity is defined as an extension of the Einstein-Hilbert action which, in the absence of matter, has the following form
\begin{equation}
    I = \frac{1}{16\pi G}\int d^4x \sqrt{-g} f(R),
\end{equation}
where $f(R)$ is an arbitrary function of the Ricci scalar. If we vary this action with respect to the metric $g_{\mu\nu}$ we obtain the vacuum field equations
\begin{equation}\label{f(R) field equations}
    f^\prime R_{\mu\nu} - \frac{1}{2}g_{\mu\nu}f - \nabla_\mu\nabla_\nu f^\prime + g_{\mu\nu}\square f^\prime = 0,
\end{equation}
where in this section we use a prime to denote the derivative with respect to $R$. Additionally, the trace of Eq. (\ref{f(R) field equations}) gives
\begin{equation}\label{trace field eq}
    \square f^\prime + \frac{Rf^\prime - 2f}{3} = 0.
\end{equation}

It is well known that the $f(R)$ gravity is equivalent to a scalar-tensor theory of gravity \cite{deFelice2010}. Therefore, we expect the same results for the polarization modes as obtained in the previous section. We show this equivalence by directly solving the equations in the weak-field approximation.

First, notice that Minkowski is not a vacuum solution of the theory. Therefore, different from Einstein's gravity, to study vacuum GWs in $f(R)$-gravity we should expand the metric around a non-flat background metric \cite{Yang2011}
\begin{equation}
    g_{\mu\nu} = g_{\mu\nu}^{\rm (b)} + h_{\mu\nu},
\end{equation}
where $g_{\mu\nu}^{\rm (b)}$ is a background metric with constant curvature (de Sitter or anti-de Sitter). In this sense, the perturbed Ricci scalar and the perturbed function $f(R)$ become
\begin{align}\label{scalar expansion}
    R & = R_{\rm b} + \delta R + {\mathcal O}(h^2),\\
    f(R) & = f(R_{\rm b}) + f^\prime(R_{\rm b})\delta R+ \frac{1}{2}f^{\prime\prime}(R_b)\delta R^2+ {\mathcal O}(h^3),
\end{align}
where $R_{\rm b}$ is the constant background curvature scalar. 

With this expansion in the Eq. (\ref{trace field eq}) we obtain 
\begin{equation}\label{trace field eq 2}
    \left(\square - m^2\right)\delta R = 0,
\end{equation}
where
\begin{equation} \label{eq mass f(R) gravity}
    m^2 \equiv \frac{1}{3}\left( \frac{f^\prime_{\rm b}}{f^{\prime\prime}_{\rm b}} - R_{\rm b}\right),
\end{equation}
and $f_{\rm b} = f(R_{\rm b})$. Notice that now all the covariant derivatives are evaluated using the background metric. 

Moreover, from the Eq. (\ref{f(R) field equations}) we obtain the following equation for the perturbation of the Ricci tensor \cite{Yang2011}
\begin{align}\label{perturbed field eq}
    \delta R_{\mu\nu}  &+ \left( \frac{f^{\prime\prime}_{\rm b}}{f^{\prime}_{\rm b}} R_{\mu\nu}^{\rm b} - \frac{1}{2} g_{\mu\nu}^{\rm b}\right)\delta R - \frac{1}{2}\frac{f_{\rm b}}{f^{\prime}_{\rm b}}\delta g_{\mu\nu} \nonumber \\ 
    &+ \frac{f^{\prime\prime}_{\rm b}}{f^{\prime}_{\rm b}} \left(g_{\mu\nu}^{\rm (b)}\square - \nabla_\mu\nabla_\nu\right) \delta R = 0.
\end{align}

At the scale size of the GW detectors, one can assume a nearly Minkowski background metric $g_{\mu\nu}^{\rm (b)} \approx \eta_{\mu\nu}$ and $R_{\mu\nu}^{\rm b} \approx 0$. Let us assume $f(R)$ models for which $f(R_{\rm b}) \approx 0$ at this limit, but in general $f_{\rm b}^\prime \neq 0$ and $f_{\rm b}^{\prime \prime} \neq 0$. This is the case of some $f(R)$ models which are viable alternatives to explain the accelerated expansion of the Universe \cite{Yang2011}. In this limit the d'Alembertian operator in Eq. (\ref{trace field eq 2}) is $\square = \eta^{\mu\nu}\partial_\mu\partial_\nu$, and Eq. (\ref{perturbed field eq}) simplifies to
\begin{equation}\label{pert field eq Mink}
 \delta G_{\mu\nu} + \frac{1}{3 m^2}\left( \eta_{\mu\nu} \square - \partial_\mu\partial_\nu\right)\delta R = 0,
\end{equation}
 which is identical to Eq. \ref{eq perturbed eq scalar tensor}. Therefore, we can follow the same procedure as in the case of scalar-tensor theories to find the equations for the gauge-invariant variables in the $f(R)$-gravity
\begin{equation}
    \left(\square - m^2\right)\Phi = 0,
\end{equation}
\begin{equation}
    \nabla^2 \Xi_i = 0,
\end{equation}
\begin{equation}\label{tensor perturbation f(R)}
    \square h_{ij}^{TT} = 0,
\end{equation}
where
\begin{equation}\label{psi phi equality}
    \Psi = \Phi = -\frac{R}{6m^2},
\end{equation}
and again we conclude that 
\begin{equation}\label{xi eq 0}
    \Xi_i = 0.
\end{equation}

Thus, as in scalar-tensor theories of gravity, we conclude that $f(R)$-gravity presents three propagating degrees of freedom. The usual two tensor modes propagating at the speed of light and one scalar degree of freedom with transversal and longitudinal behavior. Such a conclusion was found previously by Moretti et al. \cite{Moretti2019}, where the authors have also used gauge-invariant variables to describe the polarization modes, though with some slight differences when compared with our derivation. An analog discussion about the degrees of freedom and the polarization states of GWs in $f(R)$-gravity can be found, e.g., in Ref. \cite{Liang2017} where the Lorentz gauge has been used. Regarding Lorentz transformations, GWs in the scope of $f(R)$-gravity are also within case 1 of Section \ref{sec: Lorentz transf}. 

\subsubsection{A class of quadratic theories of gravity}

Now, let us apply the gauge-invariant formalism to a wide class of alternative theories of gravity considered by Yunes and Stein \cite{Yunes2011} in the scope of nonspinning black holes and the parametrized post-Einsteinian framework for GWs. The modified Einstein-Hilbert action is given by
\begin{align}\label{Eq action wide class theory}
    I = & \int d^4 x \sqrt{-g}\Big\{ \frac{R}{16\pi G} + \tilde{\alpha}_1f_1(\varphi)R^2 + \tilde{\alpha}_2 f_2(\varphi)R_{\mu\nu}R^{\mu\nu} \nonumber\\
    & + \tilde{\alpha}_3 f_3(\varphi) R_{\lambda\mu \nu \kappa}R^{\lambda\mu \nu \kappa} + \tilde{\alpha}_4f_4(\varphi){R_{\lambda\mu \nu \kappa}}^\ast R^{\lambda\mu \nu \kappa} \nonumber \\
    &- \frac{\beta}{2}[\nabla_\mu \varphi \nabla^\mu\varphi + V(\varphi)] \Big\}, 
\end{align}
where ${R_{\lambda\mu \nu \kappa}}^\ast$ is the dual of the Riemann tensor, $\varphi$ is a scalar-field, and $(\tilde{\alpha}_i,\beta)$ are coupling constants. Theories described by actions of this type can be motivated by the low-energy expansions of string theory. Notice that the above action differs from that considered in \cite{Tachinami2021}.

Since we have already discussed the role of the scalar fields in the polarization modes of GWs, let us consider only the case of constant couplings, and the scalar field is absent ($\beta = 0$). Therefore we can set $f_i(\varphi) = 1$, and from the variation of the action (\ref{Eq action wide class theory}) we obtain the field equations
\begin{equation}\label{Eq. Field eq General theory}
    G_{\mu\nu} + \alpha_1 \mathcal{H}_{\mu\nu} + \alpha_2 \mathcal{I}_{\mu\nu} + \alpha_3 \mathcal{J}_{\mu\nu} = 0,
\end{equation}
where $\alpha_i = 16\pi G\tilde{\alpha}_i$, and 
\begin{align}
    \mathcal{H}_{\mu\nu} \equiv~ & 2R_{\mu\nu} R - \frac{1}{2} g_{\mu\nu}R^2 - 2 \nabla_\mu \nabla_\nu R + 2 g_{\mu\nu} \Box R,  \\
    \mathcal{I}_ {\mu\nu} \equiv~ & \Box R_{\mu\nu} + 2 R_{\mu\lambda\nu\kappa}R^{\lambda\kappa} - \frac{1}{2}g_{\mu\nu} R_{\lambda\kappa}R^{\lambda\kappa} \nonumber \\
    &+ \frac{1}{2}g_{\mu\nu}\Box R - \nabla_\mu \nabla_\nu R,  \\
    \mathcal{J}_ {\mu\nu} \equiv ~& 8 R^{\lambda\kappa}R_{\mu\lambda\nu\kappa} - 2g_{\mu\nu}R^{\lambda\kappa}R_{\lambda\kappa} + 4\Box R_{\mu\nu} \nonumber \\
    &- 2RR_{\mu\nu} + \frac{1}{2}g_{\mu\nu}R^2 - 2\nabla_\mu\nabla_\nu R.
\end{align}

The trace of Eq. (\ref{Eq. Field eq General theory}) is 
\begin{equation}\label{eq curvature scalar quadr gravity}
    2(3\alpha_1 + \alpha_2 + \alpha_3)\Box R - R = 0.
\end{equation}

Notice that in the case of $f(R)$-gravity, a Klein-Gordon type equation was found for the first-order perturbation of the Ricci scalar [see Eq. (\ref{trace field eq 2})]. On the other hand, the above equation for quadratic gravity is valid for the Ricci scalar in general, and not only for its perturbation. Nonetheless, in the present article, we are concerned only with the first-order perturbation of the Eq. \ref{eq curvature scalar quadr gravity} considering some background metric
\begin{equation}\label{eq delta R quadr grav}
    2(3\alpha_1 + \alpha_2 + \alpha_3)\Box \delta R - \delta R = 0.
\end{equation}

To obtain the perturbation equations we can start considering a non-flat background metric, and then make the flat approximation in the region of a detector as we did for $f(R)$-gravity. However, in the present case, we consider a Minkowski background metric from the beginning for simplicity. As we saw in the previous subsection, the difference between the two approaches is a redefinition of the mass parameters of the theory (see Eq. \ref{eq mass f(R) gravity}).   

Thus, with a Minkowski background metric, the perturbation of the Eq. (\ref{Eq. Field eq General theory}) reads
\begin{align}\label{eq delta G quadr grav}
    (\alpha_2 &+ 4\alpha_3) \Box \delta G_{\mu\nu} + \delta G_{\mu\nu} \nonumber \\
    &+ (2\alpha_1 + \alpha_2 + 2\alpha_3)(\eta_{\mu\nu} \Box  - \partial_\mu \partial_\nu )\delta R = 0.
\end{align}

It is worth noticing that the above equation is a generalization of the Eq. (\ref{pert field eq Mink}) including a term with a D'Alembertian operator applied to $\delta G_{\mu\nu}$. Notice that within the present framework, it is possible to identify four subclasses of theories from quadratic gravity, not only two as shown by Tachinami et al. \cite{Tachinami2021}. In what follows we show that these subclasses can differ in the number of independent radiative degrees of freedom, polarizations, and the expressions for the scalar polarizations. 

\begin{itemize}
    \item Subclass 1: $(3\alpha_1 + \alpha_2 + \alpha_3) > 0$, $(\alpha_2 + 4\alpha_3) < 0$
\end{itemize}

In this subclass, we have the following equations from Eq. (\ref{eq delta R quadr grav}) and the 00 component of Eq. (\ref{eq delta G quadr grav})
\begin{equation}\label{eq for delta R quadr grav 2}
\left( \Box - m^2 \right) \delta R =  0,   
\end{equation}
and 
\begin{equation}\label{eq for phi quadr grav}
\left( \Box - M^2 \right) \Phi =  - \gamma \delta R,   
\end{equation}
where 
\begin{align}\label{eq masses quadr}
    m^2 \equiv & \frac{1}{2(3\alpha_1 + \alpha_2 + \alpha_3)},~~M^2 \equiv -\frac{1}{(\alpha_2 + 4\alpha_3)}, \nonumber \\
    \gamma \equiv & \frac{2\alpha_1 + \alpha_2 + 2\alpha_3}{2(\alpha_2 + 4\alpha_3)} = \frac{1}{6}\left[1 - \left( \frac{M}{m} \right)^2 \right].
\end{align}

From the $0i$ components of Eq. (\ref{eq delta G quadr grav}) and the equations for the scalars we obtain the equation for the gauge-invariant vector
\begin{equation}\label{eq vector quadr grav}
    \left( \Box - M^2 \right) \Xi_i = 0.
\end{equation}

Finally, from the $ij$ components of Eq. (\ref{eq delta G quadr grav}) we obtain
\begin{equation}\label{eq for h general yet}
    \partial_i \partial_j \left(\Box \Psi - M^2 \Psi + \gamma R\right) + \frac{1}{2}\Box \left( \Box h_{ij}^{TT} - M^2 h_{ij}^{TT} \right) = 0.
\end{equation}

Taking the divergence of the above equation and remembering that $\partial^i h_{ij}^{TT} = 0$, we obtain
\begin{equation}\label{eq for psi quadr grav}
    \left(\Box - M^2 \right) \Psi = - \gamma \delta R, 
\end{equation}
and comparing with the Eq. (\ref{eq for phi quadr grav}) we conclude that 
\begin{equation}
    \Psi = \Phi.
\end{equation}

Using this identity in the definition of the perturbed curvature scalar (\ref{curvature scalar}), the equations (\ref{eq for delta R quadr grav 2}) and (\ref{eq for phi quadr grav}) result
\begin{equation}\label{eq phi final}
    \Box \left( \Box - m^2 \right)\Phi = 0.
\end{equation}

Finally, we obtain the equations for the tensor gauge-invariant quantities by using the equation (\ref{eq for psi quadr grav}) back in Eq. (\ref{eq for h general yet}) 
\begin{equation}\label{eq for hTT final quadr grav}
    \Box \left( \Box h_{ij}^{TT} - M^2 h_{ij}^{TT} \right ) = 0.
\end{equation}

Therefore, we conclude that all the gauge-invariant variables describe the propagating modes of GWs. From Eq. (\ref{eq vector quadr grav}) we have two propagating vector degrees of freedom with mass $M$. For the scalar sector, we can obtain two independent propagating solutions from Eq. (\ref{eq phi final}), one describes a massless scalar and the other a massive scalar with mass $m$. A similar conclusion can be drawn for the tensor sector from Eq. (\ref{eq for hTT final quadr grav}). Now, we have two massless degrees of freedom and two degrees of freedom of a massive tensor field with mass $M$. Thus, the quadratic gravity in this general case presents 8 propagating degrees of freedom. 

All six polarization modes of GWs are present in this case. Only the massive solution of the scalar $\Phi$ contributes to the scalar longitudinal polarization described by the variable $\Theta$. On the other hand, both solutions contribute to the scalar transversal mode. Similarly, the massless and the massive solutions of the tensor variable describe the $+$ and $\times$ polarizations simultaneously. Finally, since $\Xi_i$ are two non-null propagating modes, none of the variables are Lorentz invariant, as seen from the Eqs. (\ref{eq Lorentz phi}), (\ref{eq Lorentz Psi}),(\ref{eq Lorentz Xi}), and (\ref{eq Lorentz h}).  

As a remark notice that, for the scalar sector the solution of a system of equations similar to Eqs. (\ref{eq for delta R quadr grav 2}) and (\ref{eq for phi quadr grav}) in the presence of matter was found by Vilhena et al. \cite{Vilhena2021}.

\begin{itemize}
    \item Subclass 2: $(2\alpha_1 + \alpha_2 + 2\alpha_3) = 0$
\end{itemize}

This is a subclass of theories whose solution can be obtained from the previous general solution. From Eq. (\ref{eq masses quadr}) notice that $\gamma = 0$ and so there is only one mass scale 
\begin{equation}
 m^2 = M^2 = \frac{1}{2(\alpha_1 - \alpha_3)}.   
\end{equation}

Using $\gamma =0$ in Eq. (\ref{eq for phi quadr grav}) and (\ref{eq for psi quadr grav}) we find that the scalars respect the same Klein-Gordon type equation as well as $\delta R$ [see Eq. (\ref{eq for delta R quadr grav 2})]. On the other hand, we can not conclude that $\Psi = \Phi$ as in the previous case, but they are related through Eq. (\ref{curvature scalar}). If we further assume that the scalars, $\Psi$, $\Phi$, and $\delta R$ are functions of the retarded time, the Eq. (\ref{curvature scalar}) leads to
\begin{equation}
    \delta R = 2\omega^2\left[ \eta^2\Psi + (2\eta^2 - 3)\Phi \right],
\end{equation}
where $\eta = \sqrt{1 - (m/\omega)^2}$.

Thus, this subclass presents only one independent scalar degree of freedom. It is a propagating mode provided $\alpha_1 > \alpha_3$. The number of degrees of freedom in the vector and tensor sectors is the same as before since they respect the same equations. We conclude that seven degrees of freedom describe the six polarization modes of GWs.

\begin{itemize}
    \item Subclass 3: $(3\alpha_1 + \alpha_2 + \alpha_3) = 0$
\end{itemize}

Within this condition, the Eq. (\ref{eq delta R quadr grav}) reduces to  
\begin{equation}\label{eq delta r again}
    \delta R = 0,
\end{equation}
and the Eq. (\ref{eq delta G quadr grav}) simplifies to
\begin{equation}
    (\alpha_2 + 4\alpha_3) \Box \delta G_{\mu\nu} + \delta G_{\mu\nu} = 0.
\end{equation}

From these equations, we obtain $\Phi$ and $\Psi$ satisfying a Klein-Gordon equation, each with the same mass 
\begin{equation}
    M^2 = \frac{1}{3(\alpha_1 - \alpha_3)}.
\end{equation}

The relation between the two scalars can be found after combining Eqs. (\ref{curvature scalar}) and (\ref{eq delta r again}). Assuming they are oscillatory functions of the retarded time, we find 
\begin{equation}
    \Psi = \left( \frac{3}{\eta^2} - 2 \right) \Phi,
\end{equation}
where $\eta = \sqrt{1 - (M/\omega)^2}$, and the scalar longitudinal polarization is described by 
\begin{equation}
    \Theta = 2(1-\eta^2)\Phi = 2 \left(\frac{M}{\omega} \right)^2 \Phi.
\end{equation}

The two vector modes also have mass $M$ and the tensor modes are described by two massless and two massive degrees of freedom with the mass $M$. Thus, again we have seven independent degrees of freedom describing the six polarization states of GWs. As before, we must have $\alpha_1 > \alpha_3$ to the massive modes represent propagating modes.

\begin{itemize}
    \item Subclass 4: $\alpha_2 = - 4\alpha_3$
\end{itemize}

In this subclass, the term with a D'Alembertian operator applied to the Einstein tensor vanishes in Eqs. (\ref{eq delta G quadr grav}) and we obtain 
\begin{equation}
    \delta G_{\mu\nu} + (2\alpha_1 + \alpha_2 + 2\alpha_3)(\eta_{\mu\nu} \Box  - \partial_\mu \partial_\nu )\delta R = 0,
\end{equation}
which is identical in form to the Eq. (\ref{linearized eqs scalar tensor}) of the scalar-tensor theory and to the Eq. (\ref{pert field eq Mink}) of $f(R)$-gravity. Therefore, we obtain the same results as in those cases. We have only one independent scalar degree of freedom, and the scalars respect the identity 
\begin{equation}
    \Psi = \Phi = - \frac{\delta R}{6 m^2},
\end{equation}
with the mass $m$ given by the definition (\ref{eq masses quadr}). In the present case, it simplifies to
\begin{equation}
    m^2 = \frac{1}{6(\alpha_1 - \alpha_3)}.
\end{equation}

Again, the scalar degree of freedom is a propagating mode provided $\alpha_1 > \alpha_3$.

The vector components vanish identically and we have only the two massless propagating degrees of freedom in the tensor sector. Theories within this subclass present four polarizations, the scalar transversal described by $\Phi$, the scalar longitudinal $\Theta = - (m/\omega)^2 \Phi$, and the two tensor polarizations propagating at the speed of light. Furthermore, the variables $\Phi$, $\Theta$, and $\Xi_i$ are Lorentz invariant quantities since this subclass of theories entered in case 1 discussed in Section \ref{sec: Lorentz transf}.

\subsubsection{Towards a general parametrization}

The linearized field equations of the theories studied in the preceding subsections are subclasses of the following equations 
\begin{align}\label{eq general linear equation}
   A_1&\Box\delta G_{\mu\nu} + \delta G_{\mu\nu} \nonumber \\
   &+ \frac{1}{3}\left(\eta_{\mu\nu}\Box - \partial_\mu \partial_\nu \right) \left[(A_1 + A_2)\delta R + 3B \frac{\delta \varphi}{\varphi_0} \right] = 0, 
\end{align}
where $A_1$ and $A_2$ are constants with dimension of square length and $B$ is a dimensionless constant. The above equation must be supplemented by
\begin{equation}
    \Box \delta \varphi =0,
\end{equation}
if there is a coupling with a massless scalar field ($B\neq 0$). On the other hand, the Eq. (\ref{eq general linear equation}) with $B = 0$ and $A_1 = 0$ is enough to describe the case of non-minimal coupling with a massive scalar field. This was shown at the end of subsection \ref{subsection ST theory}.

Thus, comparing Eq. (\ref{eq general linear equation}) with those found for scalar-tensor theory, $f(R)$-gravity and quadratic gravity [see Eqs. (\ref{linearized eqs scalar tensor}), (\ref{eq perturbed eq scalar tensor}), (\ref{pert field eq Mink}), and (\ref{eq delta G quadr grav})], we see that it is possible to reduce the problem of determining the number of radiative degrees of freedom and the polarization content of a given theory to the problem of finding three parameters, namely, $A_1$, $A_2$, and $B$. This is true if a theory of gravity has linearized field equations given by (\ref{eq general linear equation}), which encompasses a wide variety of theories as summarized in Table \ref{Tab parameters and pol}. This table shows that these three parameters are closely related to the gauge-invariant variables. Determining their values or relations between them implies a theory's polarization content and the number of independent radiative degrees of freedom.

\subsubsection{Number of radiative degrees of freedom versus number of polarization modes} 
In the previous subsections, we noticed a distinction between the number of radiative degrees of freedom and the number of polarization modes of GWs in gravity theories. Discussions and some criticisms appear in several recent works, e.g., \cite{Liang2017} and \cite{Hou2018}. In the language of \cite{Eardley1973a,Eardley1973b}, the number of polarization modes corresponds to how GWs interact with a sphere of test particles. It has nothing to do with the number of independent dynamical degrees of freedom of the linearized theory, which can be smaller or bigger than the number of polarization modes. In the case of massless scalar GWs, for instance, relative acceleration between the particles in the sphere is observed in the direction orthogonal to the wave vector $\vec{k}$. On the other hand, for a massive scalar mode, relative accelerations are also generated for particles located in the wave's propagation direction. In the latter case, we say the theory presents two scalar polarization modes, though these modes are not independent. Therefore, GWs in the $f(R)$-gravity and the scalar-tensor theories, for instance, present four polarization modes if $m\neq0$, with three independent radiative degrees of freedom. For these theories, the number of polarization modes and independent degrees of freedom agrees only if $m=0$. In the present work, we have defined a gauge-invariant variable $\Theta$ in Eq. (\ref{Theta def}) which enables an unambiguous determination of the existence of the scalar longitudinal GW mode. For other theories studied in this article, the number of radiative degrees of freedom and the number of polarization modes are depicted in Table \ref{Tab parameters and pol}. We see that the number of independent radiative degrees of freedom and the number of polarization modes of a theory depends essentially on the parameters $A_1,~A_2$, and $B$.  

\begin{table*}

    \caption{Here we show how the number of independent radiative degrees of freedom (d.o.f.) and non-null polarization modes of GWs depend on the parameters of Eq. (\ref{eq general linear equation}).} \label{Tab parameters and pol}
    \begin{ruledtabular}
        \begin{tabular}{lcccccc}
                & \# Independent     & $h_{ij}^{TT}$    &$\Xi_i$   &$\Phi$  &$\Theta$   & {\bf Examples of theories}    \\
                & radiative d.o.f. & & & & & \\
\hline
$A_1 = A_2 = B = 0$    & 2 & \checkmark       &0         &0          &0          & GR        \\
\hline
$A_1 = A_2 = 0,~B = 1$  & 3  & \checkmark  & 0  & \checkmark  & 0  & Brans-Dicke \\
\hline
$A_1 = 0,~A_2>0,~ B=0$ & 3 & \checkmark       &0         &\checkmark &\checkmark & Horndeski, $f(R)$-gravity,   \\
    & & & & & & subclass 4 of quadratic gravity   \\
\hline
$A_1<0,~A_2>0,~ B=0$    & 8 & \checkmark      &\checkmark &\checkmark &\checkmark & subclass 1 of quadratic gravity \\
\hline
$A_1 = -A_2,~A_2>0,~ B=0$         & 7 & \checkmark      &\checkmark &\checkmark &\checkmark & subclass 2 of quadratic gravity \\
\hline
$A_1<0,~A_2 = 0,~ B=0$     & 7 & \checkmark      &\checkmark &\checkmark &\checkmark & subclass 3 of quadratic gravity
\end{tabular}
    \end{ruledtabular}
    
\end{table*}

\section{Pulsar timing sensitivity}\label{sec 4 PT}

\subsection{Gauge-invariance and physical observables}
Once we have defined the polarization modes of GWs in terms of gauge-invariant variables, the sensitivities of the GW detectors are necessarily connected to such variables. This is because gauge-invariant quantities express truly physical observables. 

In the present work, we focus on the sensitivity of the pulsar timing technique to each polarization mode of GWs. Previous works consider the synchronous gauge in evaluating the pulsar timing sensitivity. However, using this gauge leaves residual gauge freedom and, apart from this, could not be appropriate to use such a gauge in all metric theories of gravity. Moreover, any GW waveform originating from a compact binary system, for instance, should be expressed using gauge-invariant variables to ensure it is a physical quantity. To evaluate the detectability of such a signal, it should be compared against a sensitivity curve which was also evaluated using the same gauge-invariant quantities.

Taking into account these aspects, we estimate the sensitivity considering the gauge-invariant variables to evaluate the pulsar timing response to the polarization modes. The elementary observable for interferometric detectors and the pulsar timing technique is the ``one-way'' fractional frequency shift $y(t) = [\nu(t) - \nu_0]/\nu_0$, where $\nu(t)$ is the frequency of an electromagnetic signal at the time of reception $t$ and $\nu_0$ is the unperturbed frequency. To achieve our goals we need a relation between $y$ and the gauge-invariant variables. However, the majority of the derivations appearing in the literature are gauge-dependent. A full gauge-invariant derivation of $y$ was obtained by Koop and Finn \cite{Koop2014}. Their derivation is quite general and includes the GW effect and all possible contributions from the background curvature (e.g., R\o{}mer delay, aberration, Shapiro time delay, and other effects appear naturally). Furthermore, no assumptions were made about the size of the detectors compared to the GW wavelength or on the dispersion relation of GWs.

For the present article, it is enough to consider the special case of a Minkowski background. If in addition, we consider that the source and the receiver of the electromagnetic signal are at rest in the same global Lorentz frame, the equation for $y$ can be written as \cite{Koop2014}
\begin{equation}\label{eq dydt}
    \frac{dy(t)}{dt} = - \int_0^{\lambda_R} \delta {R}_{0i0j}n^in^j d\lambda,
\end{equation}
where $n^i$ is the spatial unit vector in the direction of the link between the source and the receiver, and $\lambda$  is the photon's affine parameter along its unperturbed trajectory. Therefore, the GW contribution to the time derivative of the frequency shift $y$ is given by the projection of the Riemann tensor integrated along the unperturbed null geodesic linking the source and the receiver of the electromagnetic signal. Recently, B{\l}aut \cite{Blaut2019} found the same result although using a quite different approach. 

Notice that in the Koop and Finn derivation, there is no specification of the field equations of the underlying theory of gravity. The validity of their derivation lies in using the Riemann tensor as the fundamental quantity to describe the spacetime geometry and the geodesic deviation equation has the form as it appears in GR. Therefore, Eq. (\ref{eq dydt}) is valid for all four-dimensional metric theories of gravity with these properties as, for instance, those theories presented in Section \ref{sec 3 Bardeen}. Thus, this equation is appropriate for obtaining the sensitivity of interferometers and of the pulsar timing technique to the polarization modes of GWs in alternative theories of gravity in a gauge-invariant fashion.

To evaluate the one-way response let us consider the emitter of a light signal located at point 1 at a distance $L$ from the receiver. The receiver is located at point 2 at the origin of the coordinate system. The trajectory of the light signal can be parametrized as 
\begin{equation}
    t = t_2 - (L - \lambda), ~~\vec{r} = (L - \lambda) \hat{n},
\end{equation}
with $\lambda \in [0,~L]$; $t_1$ is the time of emission, $t_2 = t_1 + L$ is the time of reception and $\hat{n}$ is the unit vector pointing from 2 to 1. Within this parametrization, the retarded times are given by
\begin{equation}
    u_A = t_2 - (1 + \eta_A\mu)(L - \lambda),~~\lambda \in [0,~L],
\end{equation}
where we have defined $\mu \equiv \hat{k}\cdot \hat{n}$.

Now, changing the variable of integration to the retarded time $u_A$, the integration along the unperturbed trajectory of the light signal in Eq. (\ref{eq dydt}) can be performed. Using the Eq. (\ref{Riemann 3rd}) in (\ref{eq dydt}) we find
\begin{align}\label{eq reception at 1}
    \frac{dy}{dt} = &-\left(\frac{\mu^2}{1 + \eta_S\mu }\right)\left[\Theta^\prime(t) - \Theta^\prime\left(t - (1+\eta_S\mu)L\right)\right] \nonumber \\
    & + \left(\frac{1 - \mu^2}{1 + \eta_S\mu}\right) \left[\Phi^\prime(t ) - \Phi^\prime\left(t - (1+\eta_S\mu)L\right)\right] \nonumber \\
    & + \left(\frac{\eta_V \mu~n^i}{1 + \eta_V\mu }\right)  \left[\Xi_i^\prime(t) - \Xi_i^\prime\left(t - (1+\eta_V\mu)L\right)\right] \nonumber \\
    & + \frac{1}{2}\left(\frac{n^in^j}{1 + \eta_T\mu }\right)\left[h_{ij}^{{\rm TT}\prime}(t) - h_{ij}^{{\rm TT}\prime}\left(t - (1+\eta_T\mu)L\right)\right]. 
\end{align}

In the final expression, we have replaced the time of reception $t_2 \rightarrow t$, a prime denotes derivative with respect to the retarded time and, for simplicity, we have considered $\eta_\Psi = \eta_\Phi = \eta_S$. In the synchronous gauge and for $\eta_A = 1$, the above equation coincides with the frequency shift derived in Refs. \cite{Tinto2010,Alves2011}. 

Therefore, we have obtained an explicit relation between a physical observable (the time derivative of the frequency shift) with gauge-invariant quantities that describe the six possible polarization states of GWs. Furthermore, we have not made any hypotheses regarding the four dispersion relations except the equality of the dispersion relations of the scalar modes.  The result above applies to pulsar timing, spacecraft Doppler tracking, and ground-based and space-based interferometric GW detectors. Here we specialize in the case of the pulsar timing technique.

\subsection{Pulsar timing responses and sensitivities}

To derive the pulsar timing response, let us consider the Earth located at the origin of a Cartesian system of coordinates with unit vectors $( \hat{i}, \hat{j}, \hat{k})$ oriented in the $x$, $y$ and $z$ directions respectively. The GW wave vector is in the direction of $\hat{k}$ and the vector $L\hat{n}$ locates the Pulsar, where $\hat{n}$ is the unit vector pointing from Earth to the Pulsar. The Pulsar emits electromagnetic signals continuously which are detected on Earth. With the help of Eq. (\ref{eq reception at 1}), we can obtain the Fourier transform of the frequency shift induced by GWs on the signal emitted by the Pulsar. Using the property of the time derivative of the Fourier transform and the relation between the time $t$ and the retarded time we find  
\begin{equation}
    \tilde{y}(f) = \tilde{y}_{SL}(f) + \tilde{y}_{ST}(f) + \tilde{y}_V(f)+ \tilde{y}_T(f), 
\end{equation}
where the Fourier transforms of the induced frequency shifts due to each gauge-invariant variable are given by
\begin{align}
    \tilde{y}_{SL}(f) &\equiv -\left(\frac{\mu^2}{1+\eta_S\mu}\right)H_{SL}(f)\left[1 - e^{i2\pi f L(1+ \eta_S\mu)}\right],\\
    \tilde{y}_{ST}(f) &\equiv \left(\frac{1-\mu^2}{1+\eta_S\mu}\right)H_{ST}(f)\left[1 - e^{i2\pi f L(1+ \eta_S\mu)}\right],\\
    \tilde{y}_V(f) &\equiv \left(\frac{\eta_V \mu~n^i}{1 + \eta_V\mu }\right)\tilde{\Xi}_i(f)\left[1 - e^{i2\pi f L(1+ \eta_V\mu)}\right], \\
    \tilde{y}_T(f) &\equiv \frac{1}{2}\left(\frac{n^in^j}{1 + \eta_T\mu }\right)\tilde{h}_{ij}^{TT}(f) \left[1 - e^{i2\pi f L(1+ \eta_T\mu)}\right],
\end{align}
where $H_{SL}(f) \equiv \tilde{\Theta}(f)$ and $H_{ST}(f) \equiv \tilde{\Phi}(f)$ are the frequency-dependent wave amplitude for the scalar longitudinal and scalar transversal polarizations respectively.

For the polarization mode `$A$' we define the angular response $R_A$ of a single pulsar timing as $R^2_{A} \equiv |\tilde{y}_{A}(f)|^2/H_{A}^2(f)$. It follows
\begin{equation}\label{eq RSL}
    R_{SL}^2 = 2 \left(\frac{\mu^2}{1+ \eta_S\mu}\right)^2 \Big[1 - \cos\big(2\pi fL(1+\eta_S \mu)\big)\Big],
\end{equation}
and
\begin{equation}\label{eq RST}
    R_{ST}^2 = 2 \left(\frac{1 - \mu^2}{1+ \eta_S\mu}\right)^2 \Big[1 - \cos\big(2\pi fL(1+\eta_S \mu)\big)\Big].
\end{equation}

In the case of vector and tensor polarization modes, we assume an elliptically polarized wave and then average over the polarizations to find the response. For an elliptically polarized vector GW, we have
\begin{equation}
    \tilde{\Xi}_i(f) = H_V(f)\left(e^{i\varphi}\sin \Gamma {\epsilon}^{(1)}_i + \cos\Gamma {\epsilon}^{(2)}_i\right),
\end{equation}
where $H_V(f)$ is the vector wave amplitude, and ${\epsilon}^{(1)}_i$ and ${\epsilon}^{(2)}_i$ are two orthogonal unit polarization vectors and both are orthogonal to $\hat{k}$. We use the polarization angles $(\varphi, \Gamma)$ to characterize elliptically polarized waves. Two particular cases are linearly and circularly polarized waves. The former can be obtained by choosing $\varphi=0$ representing a vector wave $\mathbf{\Xi}$ linearly polarized making an angle $\Gamma$ with $\hat{\bm \epsilon}^{(2)}$. Circularly polarized waves are obtained by using $\Gamma = \frac{\pi}{4}$ (the vector wave has the same amplitude in both directions) and $\varphi = \pm \frac{\pi}{2}$ (right circularly polarized wave for the plus sign and left circularly polarized wave for the minus sign). The polarization of a GW depends essentially on the generating mechanism. In what follows, we evaluate the response as an average over the polarization angles in the intervals $\varphi \in [0,~2\pi]$ and $\Gamma \in [0,~\pi]$.

An analogous expression can be written for the usual elliptically polarized tensor GWs
\begin{equation}
    \tilde{h}_{ij}^{TT}(f) =  H_T(f)\left(e^{i\varphi}\sin \Gamma {\varepsilon}^{+}_{ij} + \cos\Gamma {\varepsilon}^{\times}_{ij}\right),
\end{equation}
where we can use the pair of orthogonal vectors $(\hat{\bm \epsilon}^{(1)},\hat{\bm \epsilon}^{(2)})$ to define the two polarization tensors
\begin{align}
    {\varepsilon}^{+}_{ij} &= {\epsilon}^{(1)}_i{\epsilon}^{(1)}_j - {\epsilon}^{(2)}_i{\epsilon}^{(2)}_j,\\
    {\varepsilon}^{\times}_{ij} &= {\epsilon}^{(1)}_i{\epsilon}^{(2)}_j + {\epsilon}^{(2)}_i{\epsilon}^{(1)}_j.
\end{align}

In our Cartesian coordinate system we chose $\hat{\bm \epsilon}^{(1)}$ and $\hat{\bm \epsilon}^{(2)}$ to coincide with $\hat{i}$ and $\hat{j}$ respectively. Moreover, let us consider the usual spherical coordinates $(\theta, \phi)$ associated with the vector $L\hat{n}$ that locates the Pulsar. Notice that in this coordinate system $\mu = \hat{n}\cdot \hat{k} = \cos\theta$. Then, in the case of vector and tensor waves, we can perform an average over the polarization angles $(\varphi, \Gamma)$ to find the angular pulsar timing response 
\begin{align}\label{eq RV}
    R_V^2 &= \frac{1}{3}\left(\frac{\eta_V\mu\sqrt{1-\mu^2}}{1+ \eta_V\mu}\right)^2(1+\cos^2\phi)\nonumber\\
    &\times \Big[1 - \cos\big(2\pi fL(1+\eta_V \mu)\big)\Big],
\end{align}
and
\begin{align}\label{eq RT}
    R_T^2 &= \frac{1}{12}\left(\frac{1 - \mu^2}{1+ \eta_T\mu}\right)^2(1+\cos^22\phi)\nonumber\\
    &\times \Big[1 - \cos\big(2\pi fL(1+\eta_T \mu)\big)\Big],
\end{align}
where we have included a factor $1/2$ in both expressions since there are two vector polarizations in the first case and two tensor polarizations in the second case. 

In the present article we are interested in the sensitivity to single-source GW signals averaged over the sky and polarization states. The most promising GW sources in the pulsar timing band are supermassive binary black holes (with masses in the range $10^7 - 10^{10}~{\rm M}_\odot$) hosted in the center of galaxies. The sensitivity is defined by $\sqrt{S_y(f)B}/R^{\rm rms}_A$, where $S_y(f)$ is the one-sided power spectral density of the noise affecting the relative frequency shift of pulsar timing, and $B$ is the bandwidth. Here, we assume $B$ corresponding to an integration time of 10 years ($B = 1 {\rm cycle}/10~{\rm years}$), and a signal-to-noise ratio SNR = 1. The quantity $R^{\rm rms}_A$ is the rms of the response evaluated by performing an average over sources uniformly distributed over the celestial sphere. Thus, we find the following formula for the pulsar timing rms response 
\begin{align}\label{final rms response}
    R^{\rm rms}_A(f) = \frac{2\pi fL}{\sqrt{\eta_A}} &\left[ \sum_{n = 1}^5  \frac{a_n^A(f)}{(2\pi fL)^n} I^A_n(f)\right]^{1/2}, 
\end{align}
where, for a given polarization $A$, $I_n^A(f)$ is a set of five elementary integrals
\begin{equation}
    I_n^A(f) = - \int_{2\pi f L(1 - \eta_A)}^{2\pi f L(1 + \eta_A)} x^{n-3}(\cos x - 1)dx,
\end{equation}
where $n = 1, 2, 3, 4, 5$. Therefore, the rms of the responses differ only in the dispersion relations $\eta_A$ and in the coefficients $a_n^A$ given in the Appendix. The latter can be functions of the frequency if the speed of propagation of GWs is different from the speed of light except in the case of the scalar longitudinal polarization for which the coefficients $a_n^{SL}$ are independent of frequency for any speed.

\begin{figure}[]
    \centering
    \includegraphics[width=8cm]{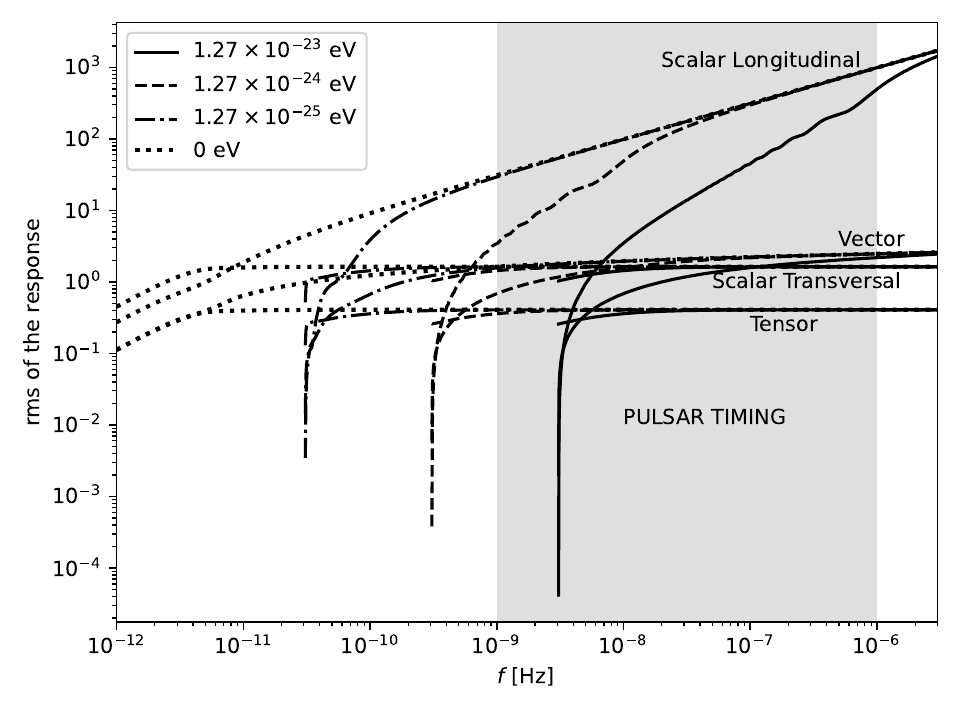}
    \caption{We show the rms of the pulsar timing response for all the gauge-invariant variables of the polarization modes. For all cases, we have used the massive dispersion relation $\eta_A(f) = \sqrt{1 - (m_A/2\pi f)^2}$. The shaded region indicates the frequency range of the pulsar timing technique. Notice that if the mass is about that of the upper bound of the LIGO detector ($1.27 \times 10^{-23}~{\rm eV}$) we have remarkable effects in this range.}
    \label{fig:PTrmsresp}
\end{figure}

Notice that in evaluating the response we have not considered any specific form for the dispersion relation. Therefore, the analytical expression (\ref{final rms response}) is a general result. To evaluate the effect of the dispersion relation on the GW response, henceforth we consider that each mode $A$ has an effective mass $m_A$ which results in $\eta_A(f) = \sqrt{1 - (m_A/2\pi f)^2}$. This dispersion relation is valid for a wide range of metric theories of gravity as we have shown in Section \ref{subsec: polarizations in alternatives}. As we have verified, the masses $m_A$ depend on the specific parameters of each theory. In Fig. \ref{fig:PTrmsresp} we show the pulsar timing rms response for the scalar longitudinal, scalar transversal, vector, and tensor polarization modes for a typical pulsar distance $L = 1~{\rm kpc}$.

 Our estimated sensitivity is based on the noise model discussed in \cite{Jenet2011}. It is assumed that timing fluctuations due to intergalactic and interplanetary plasma can be adequately calibrated,  the intrinsic pulsar rotational noise is negligible, and the pulse profile is stable. Under these assumptions, the spectrum of the noise is given by      
\begin{equation}
    S_y(f) = [4.0\times 10^{-31}f^{-1} + 3.41\times 10^{-8} f^2] {\rm Hz}^{-1},
\end{equation}
where the lower part of the frequency band ($f<3\times 10^{-8}$ Hz) is limited by the ground clock noise. For higher frequencies, the dominant noise is a white timing noise due to an uncertainty of $100$ nsec in the time of arrival of a pulse.

The resulting pulsar timing sensitivities to the polarization states are shown in Fig. \ref{fig:Sensitivity}. It shows the strength of a sinusoidal gravitational wave required to achieve a SNR = 1 over an integration time of 10 years.

\subsection{Interpretation of the sensitivities}

Notice that the sensitivity to the scalar longitudinal mode is some orders of magnitude better than the sensitivities of other polarizations, and the sensitivity to the vector modes can be up to five times better than that of the tensor mode. The response decreases as the wavelength of the GWs is of the order or larger than the distance from Earth to the Pulsar (long-wavelength limit). If $m_A = 0$ this happens for a tiny frequency, far beyond the pulsar timing frequency band $(10^{-9} - 10^{-6}~{\rm Hz})$ (see Fig. \ref{fig:PTrmsresp}). On the other hand, for a non-null mass, a fast decrease in the response can occur in this band as the technique approaches the long-wavelength limit. The cutoff frequency for which the response vanishes is related to the mass by
\begin{equation}\label{cutoff freq}
    f_c = \left(\frac{m}{m_{\rm up}}\right) 3.07 \times 10^{-9}~{\rm Hz},
\end{equation}
where we have considered the upper bound on the graviton mass imposed by LIGO, $m_{\rm up} = 1.27 \times 10^{-23}~{\rm eV}/c^2$ \cite{LIGO2022}, as a fiducial mass. Obviously, the effective mass of the vector and scalar polarizations do not need to respect this upper bound since it was derived from detections of the tensor modes.  

\begin{figure*}
     \centering
     \begin{subfigure}[b]{0.45\textwidth}
         \centering
         \includegraphics[width=\textwidth]{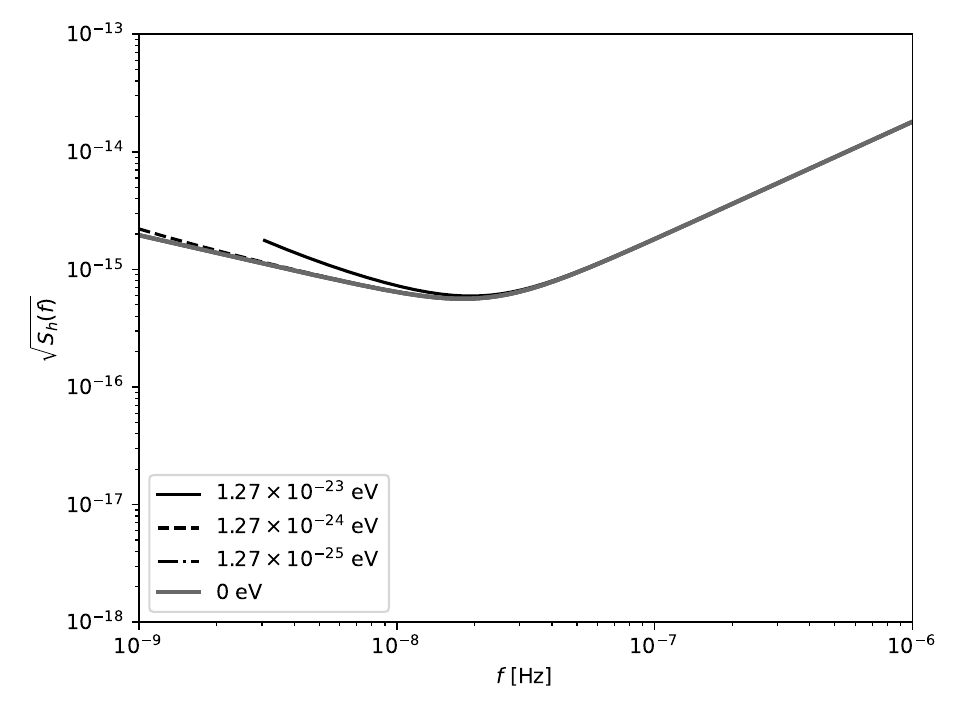}
         \caption{Tensor}
         \label{fig:tensorsens}
     \end{subfigure}
     \hfill
     \begin{subfigure}[b]{0.45\textwidth}
         \centering
         \includegraphics[width=\textwidth]{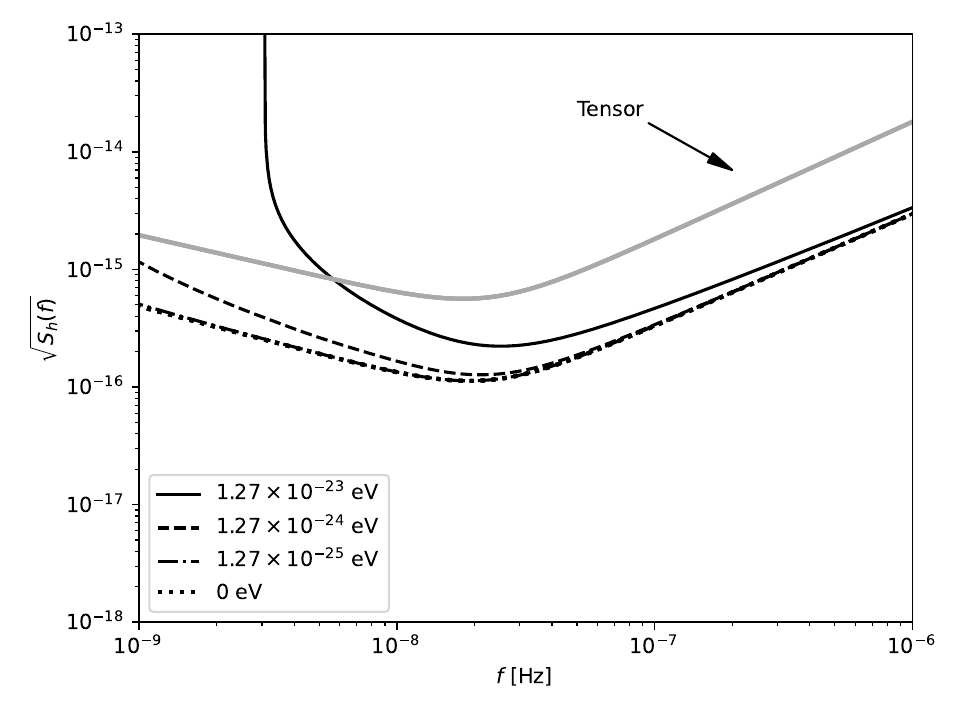}
         \caption{Vector}
         \label{fig:vectorsens}
     \end{subfigure}
     \hfill
     \begin{subfigure}[b]{0.45\textwidth}
         \centering
         \includegraphics[width=\textwidth]{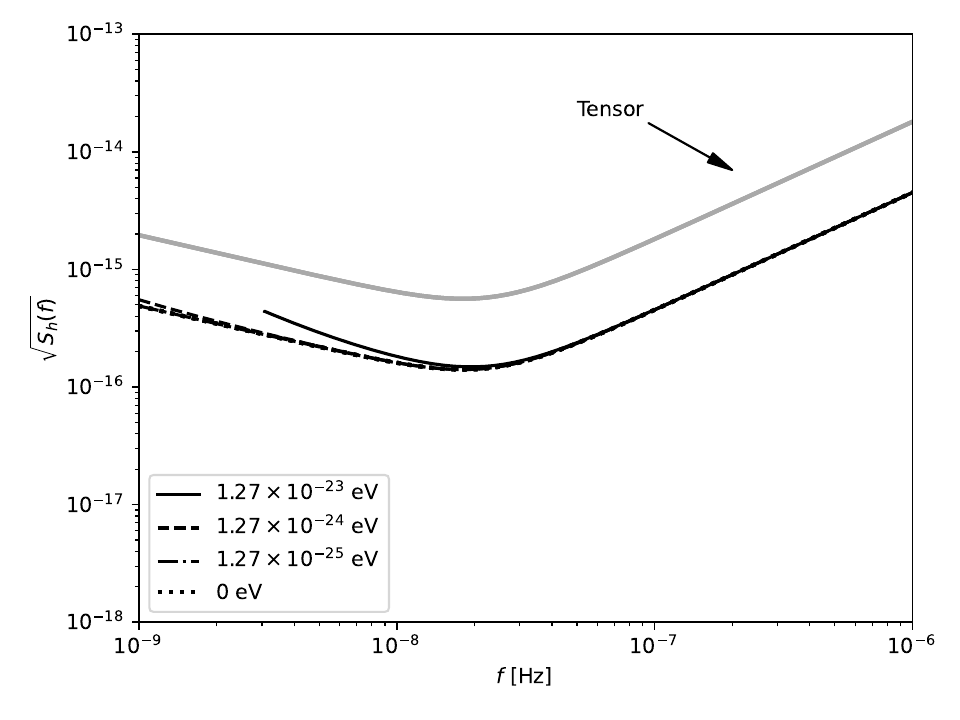}
         \caption{Scalar transversal}
         \label{fig:scalarTsens}
     \end{subfigure}
     \hfill
     \begin{subfigure}[b]{0.45\textwidth}
         \centering
         \includegraphics[width=\textwidth]{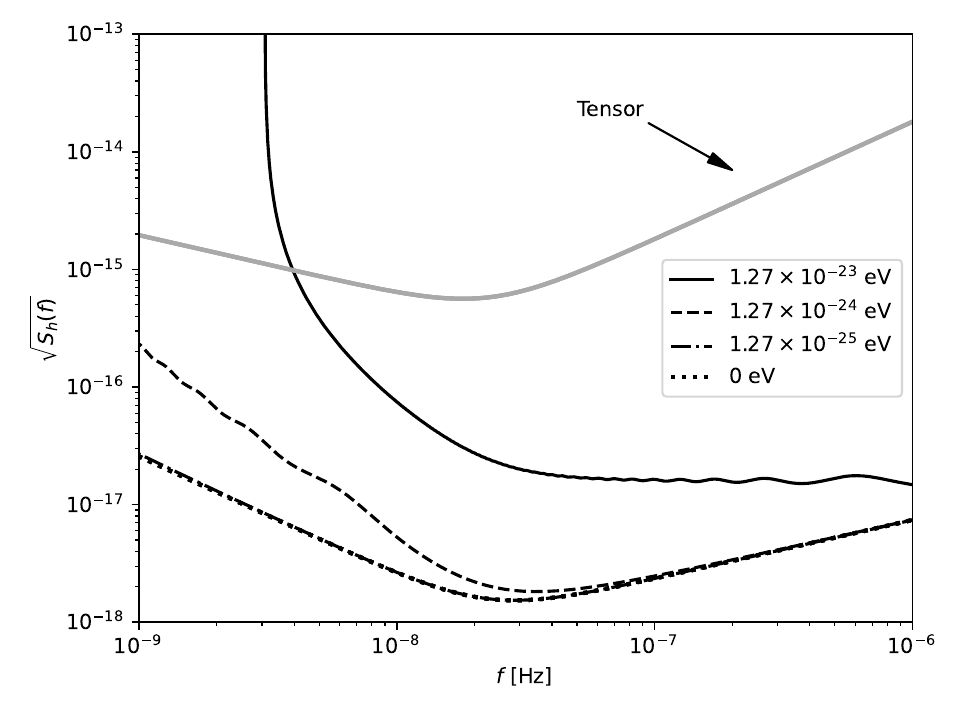}
         \caption{Scalar longitudinal}
         \label{fig:scalarLsens}
     \end{subfigure}
      
        \caption{Sensitivity curves expressed as the strength of a sinusoidal gravitational wave required to achieve a SNR = 1 over an integration time of 10 years. The sensitivities are expressed in terms of the square root of the power spectral density (PSD). It is assumed that the pulsar is at a distance of 1 kpc from Earth. We show the effect of the dispersion relation of massive GWs. The sensitivity to the massless tensor mode is shown in all figures (gray curve). Notice that as the mass approaches the upper bound of the LIGO detector ($1.27 \times 10^{-23}~{\rm eV}$) we have a remarkable change in the shape of the sensitivity curves mainly for vector and scalar longitudinal polarizations. If the mass is of this order, the cutoff frequency $f_c$ is in the pulsar timing band [see Eq. (\ref{cutoff freq})]. For lower frequencies, GWs can not be detected. Therefore, the evidence of a cutoff frequency or even the evidence that such a cutoff is not on the pulsar timing frequency band can lead to a more stringent bound of the effective mass of the graviton than that presented by ground-based interferometers.}
        \label{fig:Sensitivity}
\end{figure*}

\begin{figure*}
     \centering
     \begin{subfigure}[b]{0.48\textwidth}
         \centering
         \includegraphics[width=\textwidth]{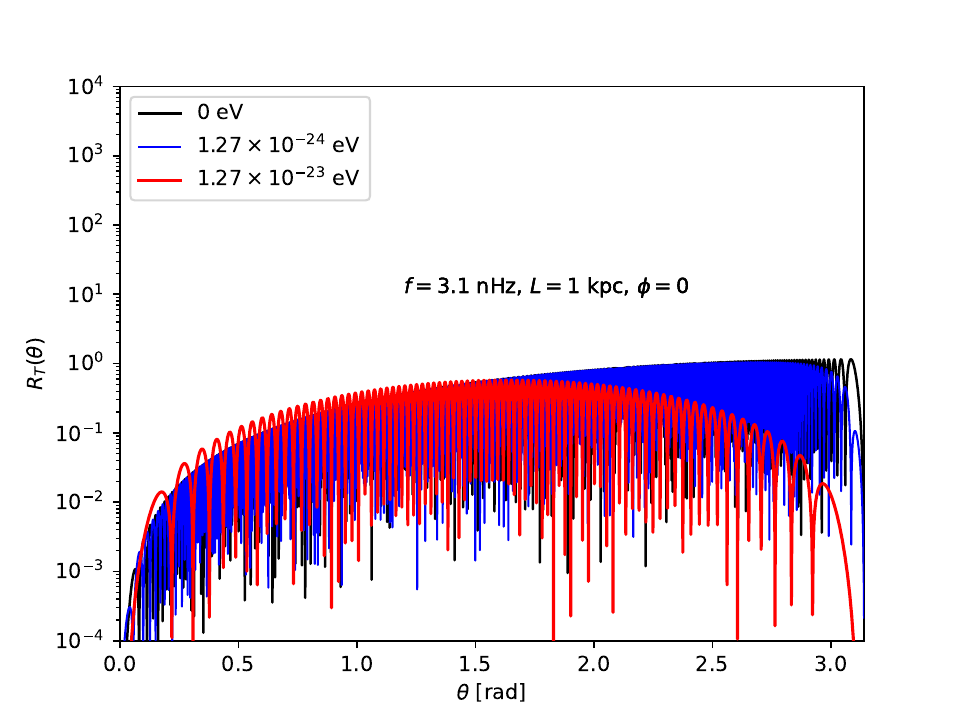}
         \caption{Tensor}
         \label{fig:tensorresp}
     \end{subfigure}
     \hfill
     \begin{subfigure}[b]{0.48\textwidth}
         \centering
         \includegraphics[width=\textwidth]{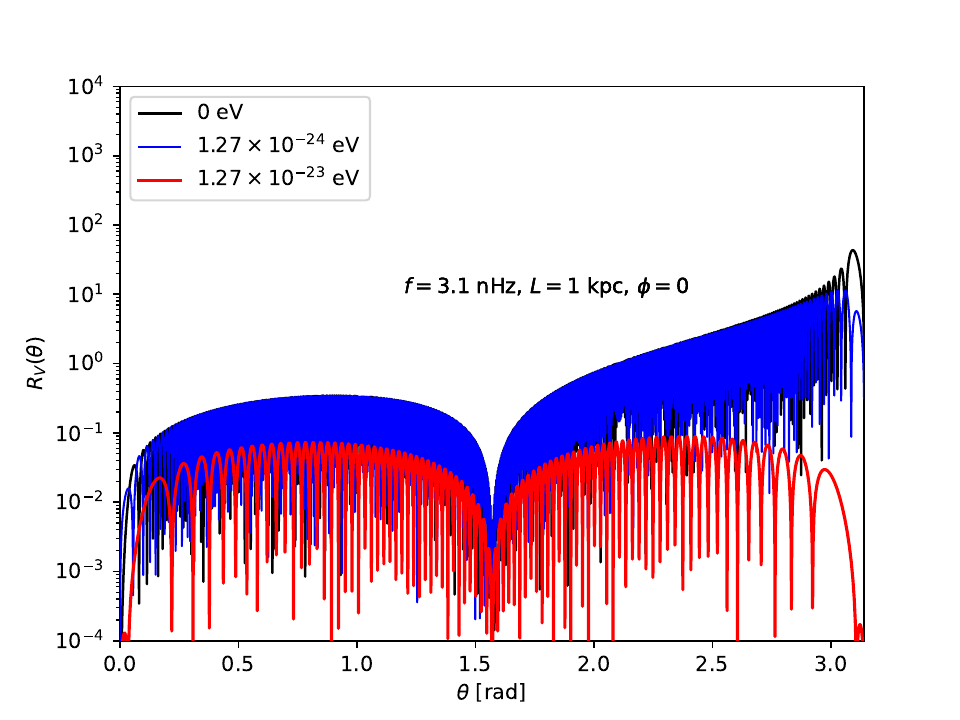}
         \caption{Vector}
         \label{fig:vectorresp}
     \end{subfigure}
     \hfill
     \begin{subfigure}[b]{0.48\textwidth}
         \centering
         \includegraphics[width=\textwidth]{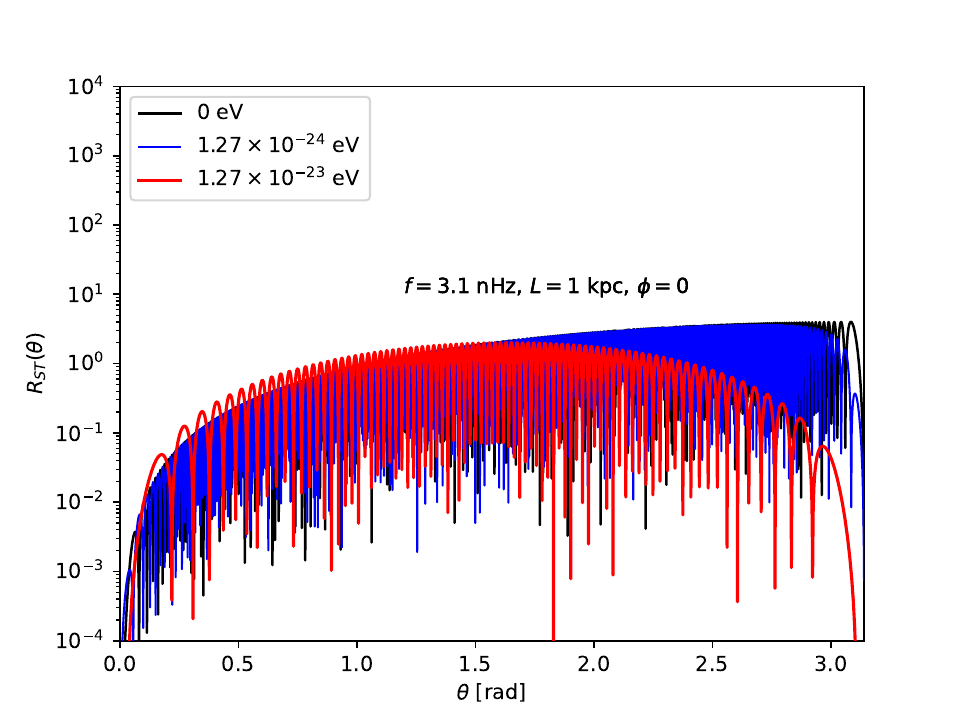}
         \caption{Scalar transversal}
         \label{fig:scalartresp}
     \end{subfigure}
     \hfill
     \begin{subfigure}[b]{0.48\textwidth}
         \centering
         \includegraphics[width=\textwidth]{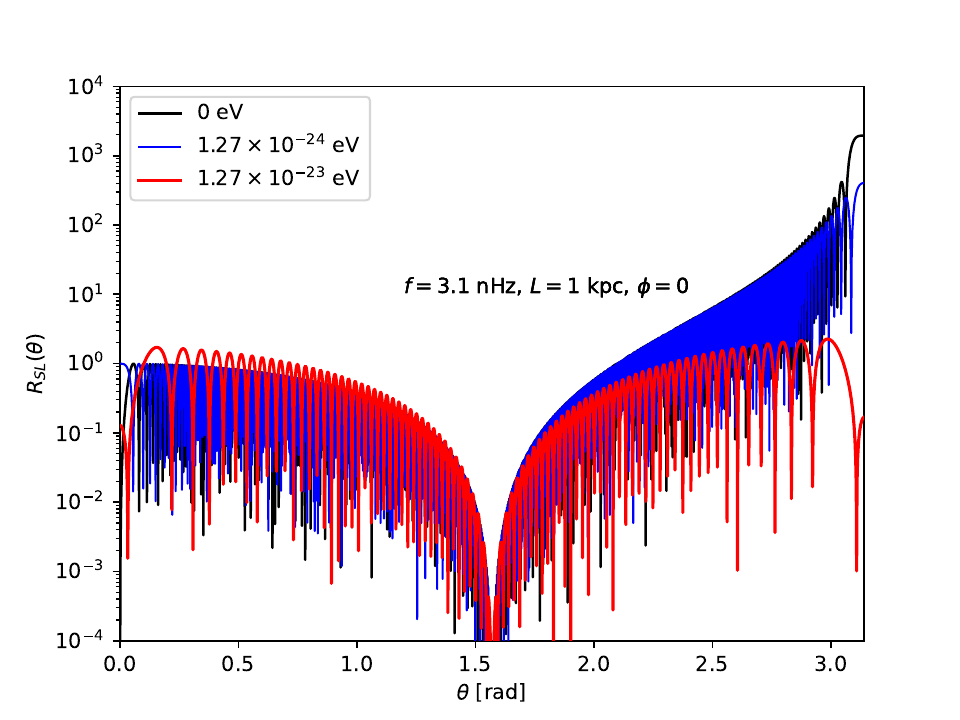}
         \caption{Scalar longitudinal}
         \label{fig:scalarlresp}
     \end{subfigure}
      
        \caption{The pulsar timing angular response for the tensor, vector, scalar transversal and scalar longitudinal polarization modes using a massive dispersion relation $\eta_A(f) = \sqrt{1 - (m_A/2\pi f)^2}$ for each case. We have considered a typical distance of $L = 1$ kpc and an angle $\phi = 0$.}
        \label{fig:three responses}
\end{figure*}

In Fig. \ref{fig:three responses} we show, as an example, the angular response (i.e., the frequency-dependent antenna pattern) given by Eqs. (\ref{eq RSL}), (\ref{eq RST}), (\ref{eq RV}) and (\ref{eq RT}) at the frequency $f = 3.1~{\rm nHz}$ and $\phi =0$ considering a single Pulsar.  We have chosen this frequency as example for two reasons. First because supermassive binary black holes have higher GW strain for lower frequencies. The second reason is that this frequency is higher but close to the cutoff frequency obtained for $m=m_{\rm up}$. Therefore, if the mass has this value, GWs with a frequency of $f = 3.1~{\rm nHz}$, for instance, approach the long-wavelength regime.

When the GW approaches the long-wavelength regime, the symmetry of the response around $\theta = \pi/2$ is restored for all the polarization modes. This case is shown in red in Fig. \ref{fig:three responses} for each polarization. The behavior of the response, in this case, is similar to that of ground-based interferometers. At $\theta = \pi/2$ the tensor polarization has the maximum response and the response vanishes for the scalar longitudinal and vector modes. On the other hand, the response for the scalar transversal mode is identical in form to the response for tensor polarization. In the same figure, we notice the oscillations in the response which comes from the square brackets in the Eqs. (\ref{eq RSL}), (\ref{eq RST}), (\ref{eq RV}) and (\ref{eq RT}). In the present case, the angles for which this term vanishes for a given frequency $f$ are given by
\begin{equation}
    \cos \theta_n = \frac{1}{\eta_A}\left(\frac{n}{fL} - 1\right),~~n = 0, 1, 2, \dots
\end{equation}

As we mentioned earlier, for the massless case the pulsar timing is out of the long-wavelength regime for the entire frequency range. In this case, we can notice an asymmetry of the response of GWs propagating in the parallel directions of the electromagnetic signal ($\pi/2 < \theta < \pi$) with respect to GWs propagating in the antiparallel directions ($0 < \theta < \pi/2$). For GWs traveling in parallel directions, the response can be some orders of magnitude higher than those traveling in antiparallel directions. This effect occurs for tensor, vector, and scalar modes. However, for the scalar longitudinal and vector modes, one can notice a remarkable enhancement of the response. This enhancement effect has been noticed for the first time by the present author and a collaborator \cite{Tinto2010,Alves2011}. It is associated with the longitudinal behavior of the mentioned polarization modes and with the relative direction of the GW wave vector $\vec{k}$ with respect to the direction of propagation of the electromagnetic signal emitted by the Pulsar.

In the present scope, the physical origin of the enhancement effect in the response of longitudinal polarizations can be understood in light of Eq. (\ref{eq dydt}). First of all, remember that in this case, the Riemann curvature tensor has components not only transverse to the direction of the propagation of the GW, but also in the longitudinal direction. Since the Riemann tensor is a function of the retarded time $u$, and $u$ depends on $\theta$, light rays coming from different directions `see' the curvature generated by the GW differently. Consider that we are out of the long-wavelength regime. The light rays traveling in the opposite directions of the GWs pass through several maxima and minima of the curvature, which makes their frequency change continuously. Since the final frequency shift measured at Earth is an integrated effect of the curvature, the result can be zero for some directions. On the other hand, those light rays propagating parallel or almost parallel to the GW experience fewer oscillations of the curvature. In this case, the final effect can be a higher frequency shift when compared with the anti-parallel case. This is because the average curvature is higher generating an increase in the response as $\theta \rightarrow \pi$. When one approaches the long-wavelength regime, the light signals originating from different directions experience fewer curvature oscillations and the curvature effect in the frequency shift becomes symmetric. Finally, in the long-wavelength regime, the curvature oscillations cannot be noticed at all in a one-way light travel. In this situation, we have the usual frequency-independent antenna patterns of ground-based interferometers.  

The same argument applies in explaining the asymmetry of the transversal polarizations (scalar transversal and tensor) out of the long-wavelength regime. But in this case, the curvature goes to zero as one approaches $\theta = 0$ or $\theta = \pi$ suppressing the enhancement effect for $\theta \rightarrow \pi$.  

In the Fig. \ref{fig:Sensitivity}, we notice that the graviton mass has a remarkable effect on the sensitivity curves as it approaches the upper bound of the LIGO detector ($1.27 \times 10^{-23}~{\rm eV}$). For tensor and scalar transversal polarizations, the predominant effect is a limit in the sensitivity established by the cutoff frequency $f_c$ given by the relation (\ref{cutoff freq}). On the other hand, for vector and scalar longitudinal polarization modes, we have a significant change in the shape of the sensitivity curve including a change in the frequency of maximum sensitivity. The sensitivity curves for massive gravitons are indistinguishable from that of the massless case if the effective mass is two orders of magnitude smaller than that of the LIGO upper bound in the case of vector and scalar longitudinal polarizations. Whereas for the transversal polarizations, it is enough that the graviton mass is one order of magnitude smaller than $m_{\rm up}$. Remember that $m_{\rm up}$ was obtained from observations of the tensor mode. This means that, in principle, the effective mass of the vector and scalar polarizations can be greater than $m_{\rm up}$. If $m$ is about three orders of magnitude higher than $m_{\rm up}$ these polarizations would be undetectable in the pulsar timing frequency band.      

\section{Conclusion}\label{sec 5 conclusion}

We have shown that the Bardeen framework enables a clear description of the six polarization modes of GWs even if each mode has a general dispersion relation. The response given by Eq. (\ref{eq reception at 1}) shows an explicit relation between a physical observable (the derivative of the frequency shift) and the gauge-invariant variables. Therefore, this relation means we have a bridge between theory and experiment, avoiding possible ambiguities of gauge choice. A new gauge-invariant variable was introduced [see Eq. (\ref{Theta def})] aiming for an unambiguous description of the scalar longitudinal polarization mode. 

In the case of a single pulsar timing, we obtained an analytical formula for the rms response [see Eq. (\ref{final rms response})] which is valid for any dispersion relation. In the case of a dispersion relation of a massive particle, we have seen that it has a significant impact on the pulsar timing sensitivity to scalar longitudinal and vector GWs. Remarkably, the effects of the mass on the pulsar timing sensitivity are particularly noticeable if it is of the order of the LIGO's upper bound for the graviton mass ($m_{\rm  up}$). If the mass is two orders of magnitude smaller than $m_{\rm up}$, the sensitivity curves are indistinguishable from the massless case. On the other hand, in the case of the scalar transversal and the tensor polarization modes, it is enough that the mass is one order of magnitude smaller than $m_{\rm up}$ to disregard its effects on the sensitivity. With a dispersion relation of massive particles the main physical effect in the case of pure transversal modes is a limitation in the detectability of these modes established by a cutoff frequency that depends on the mass. Notice that the effects on the sensitivity appear in the case of pulsar timing because the cutoff frequency we have considered lies in the pulsar timing frequency band. But, in principle, the cutoff frequency can be higher than the pulsar timing band in the case of vector and scalar polarizations. If this happens, such modes would be undetectable by pulsar timing experiments. In other words, the absence of detection does not imply that extra polarization states beyond the tensor polarization do not exist. In the future, we plan to analyze other dispersion relations of GWs appearing in the literature to check their implications on the pulsar timing sensitivity.

The detection (or absence of detection) of the polarization modes using the pulsar timing technique has decisive implications for alternative theories of gravity. Consider, for instance, the case of the theories studied in Section \ref{subsec: polarizations in alternatives} for which the tensor mode is massless and the scalar modes can be massive. Suppose that the scalar mode has a mass of about that of the LIGO upper bound, therefore for frequencies approaching the cutoff $f_c \sim 3 \times 10^{-9}$ Hz the sensitivity of the scalar longitudinal polarization becomes worse than that of the tensor modes. Below this frequency, the scalar modes could not be detected (or even be produced!). Thus, suppose we are looking for GWs only in a frequency band below $f_c$, and we detect only tensor polarizations. We could be led to the wrong conclusion that the scalar modes do not exist. On the other hand, if there is a cutoff frequency for the scalar modes, but not for the tensor modes, this could corroborate the scalar-tensor theories of gravity or $f(R)$-gravity. Moreover, this would lead to a bound on the mass of the scalar mode.  

We have seen that the pulsar timing sensitivity to the scalar longitudinal mode is some orders of magnitude better than the sensitivity to tensor modes. However, depending on the theory of gravity this could not be an advantage for detecting this mode. In the case of the theories we have analyzed, the amplitude of the scalar longitudinal mode is related to the amplitude of the scalar transversal mode through a factor $(m/\omega)^2$ [see, for instance, Eq. (\ref{eq: relation theta and mass})]. Therefore, if $m$ is much smaller than the smallest detectable frequency of pulsar timing, the scalar-longitudinal mode can become undetectable even if the scalar transversal mode is detected. Obviously, these results apply to scalar-tensor theories of gravity,  to $f(R)$-gravity and also to some subclasses of the quadratic gravity.  Other theories may have a different relation between $\Theta$, $\Phi$, the mass, and the frequency, leading to different conclusions. 

Our analysis shows that the evidence of a cutoff frequency for any polarization or that such a cutoff is not in the pulsar timing band can lead to a more stringent bound on the graviton mass than that presented by ground-based interferometers.

Pulsar timing detection presents a great opportunity to test gravity by imposing bounds on the polarization modes of GWs. However, to impose such bounds, it is necessary to compute the expected GW strain in the pulsar timing band. The most promising source of GWs in the frequency band $10^{-9} - 10^{-6}~{\rm Hz}$ is supermassive binary black holes with masses in the range $10^7 - 10^{10}~{\rm M}_\odot$. This could be individual sources or an incoherent superposition of the cosmic population of such systems forming a stochastic background. The GW strains corresponding to each polarization mode generated by supermassive black holes depend on the details of each specific theory. The evaluation of them is out of the scope of the present article since we are interested in a general formalism for both describing the polarization modes and their corresponding pulsar timing sensitivity.

Finally, to evaluate the sensitivity to a stochastic background of astrophysical or cosmological origin, it is necessary to consider an array of pulsars and the correlation functions between them, that are distinct for each polarization mode. The derivation of such sensitivity with gauge-invariant variables is the subject of a forthcoming publication.   

\begin{acknowledgments}
The author thanks Dr. Massimo Tinto for helpful discussions and encouragement, and Lívia R. Alves for continuous encouragement during the development of this work. 
\end{acknowledgments}

\section*{Appendix}

Here we give the frequency-dependent quantities that appear in Eq. (\ref{final rms response}). 

For the scalar-longitudinal response
\begin{align}
    a^{SL}_1 & = 1 , \\
    a^{SL}_2 & = -4, \\
    a^{SL}_3 & = 6, \\
    a^{SL}_4 & = -4, \\
    a^{SL}_5 & = 1.
\end{align}

For the scalar-transversal response 
\begin{align}
    a^{ST}_1 & = \left[1 - \frac{1}{\eta_{S}^2(f)}\right]^2 , \\
    a^{ST}_2 & = \frac{4}{\eta_{S}^2(f)}\left[1 - \frac{1}{\eta_{S}^2(f)}\right], \\
    a^{ST}_3 & = \frac{2}{\eta_{S}^2(f)}\left[\frac{3}{\eta_{S}^2(f)} - 1\right], \\
    a^{ST}_4 & = -\frac{4}{\eta_{S}^4(f)}, \\
    a^{ST}_5 & = \frac{1}{\eta_{S}^4(f)}.
\end{align}

For the vector response
\begin{align}
    a^V_1 & = \frac{1}{4} \left[1 - \frac{1}{\eta_V^2(f)}\right], \\  
    a^V_2 & = \frac{1}{2}\left[\frac{2}{\eta_V^2(f)}  - 1\right], \\  
    a^V_3 & = \frac{1}{4}\left[ 1 - \frac{6}{\eta_V^2(f)} \right], \\  
    a^V_4 & = \frac{1}{\eta_V^2(f)}, \\  
    a^V_5 & = - \frac{1}{4\eta_V^2(f)} . 
\end{align}

For the tensor response
\begin{align}
    a^T_1 & = \frac{1}{16}\left[ 1 - \frac{1}{\eta_T^2(f)}\right]^2,\\
    a^T_2 & = \frac{1}{4\eta_T^2(f)} \left[ 1 - \frac{1}{\eta_T^2(f)}\right], \\
    a^T_3 & = \frac{1}{8\eta_T^2(f)} \left[\frac{3}{\eta_T^2(f)} - 1\right] \\
    a^T_4 & = - \frac{1}{4\eta_T^4(f)}, \\
    a^T_5 & = \frac{1}{16 \eta_T^4(f)}.
\end{align}


\section*{References}
\bibliography{article}

\end{document}